\documentclass{article}

\usepackage{latexsym}
\usepackage{amssymb}
\usepackage{amsmath}
\usepackage{dsfont}
\usepackage{graphicx}
\usepackage{subcaption}

\newcommand{\ket}[1]{\left| #1 \right\rangle}
\newcommand{\bra}[1]{\left\langle #1 \right|}
\newcommand{\braket}[2]{\langle #1 | #2 \rangle}

\newcommand{\tr}{\mathrm{tr}}
\newcommand{\comment}[1]{}
\newcommand{\cor}{\zeta}

\newcommand{\grad}{\nabla}

\newtheorem{thm}{Theorem}
\newtheorem{corollary}[thm]{Corollary}
\newtheorem{proposition}[thm]{Proposition}
\newtheorem{lemma}[thm]{Lemma}
\newtheorem{definition}[thm]{Definition}
\newtheorem{example}[thm]{Example}
\newtheorem{remark}[thm]{Remark}
\newtheorem{assumption}[thm]{Assumption}

\begin{document}

\title{An analysis of the stationary operation of atomic clocks}
\author{Martin Fraas \\ \small{2521 San Pablo Ave., Berkeley, CA 94702}}

\maketitle

\begin{abstract}
We develop an abstract model of atomic clocks that fully describes the dynamics of repeated synchronization between a classical oscillator and a quantum reference. We prove existence of a stationary state of the model and study its dependence on the control scheme, the interrogation time and the stability of the oscillator. For unbiased atomic clocks, we derive a fundamental bound on atomic clocks long time stability for a given local oscillator noise. In particular, we show that for a local oscillator noise with integrated frequency variance scaling as $T^\alpha$ for short times $T$, the optimal clock time variance scales as $F^{-(\alpha +1)/(\alpha +2)}$ with respect to the quantum Fisher information, $F$, associated to the quantum reference.

In an attempt to prove the bounds without the unbiasedness assumption, we derive a new Cramer-Rao type inequality.
\end{abstract}

\section{Introduction}

The significance of Atomic clocks is evident in many contemporary scientific and technological endeavors. Perhaps most fundamental is the fact that they underlie our very notion of time; The SI unit of time is defined in terms of a transition in the Cesium atom, and a network of primary frequency standards, which are atomic clocks based on this transition, form the International Atomic Time. Equally significant, and maybe practically more important, is that atomic clocks are an essential building block in widely used technological systems such as satellites and communication devices. 

Atomic clocks reached an unprecedented low frequency uncertainty. In a recent table top experiment \cite{Chou}, atomic clocks were used to measure the general relativity metric of earth. In this experiment, a relative frequency difference between two optical clocks is measured before and after one of the clocks is lifted by $50 cm$. This reveals a shift in the relative frequency difference of order $10^{-17}$ that matches the general relativity prediction. Besides demonstrating atomic clocks accuracy, the experiment may lead to new applications of atomic clocks in geodesy \cite{Kleppner}.

Naturally, we are in the pursuit of even better atomic clocks. Although in the past years, the progress in their construction is mainly driven by advances in laser spectroscopy and atom manipulation techniques, the pursuit also brings the question if there are ultimate limits on how good clocks we can built. As far as the author knows, the question have been first studied by Wigner \cite{Wigner}, but the theoretical problem of measurement of time in general goes back to Poincare \cite{Poincare}. The atomic clock theory brings these questions to a very concrete level.

We outline the challenges in the atomic clock theory in the following section, which, however, starts with an explanation of the atomic clocks' operation. This is followed by a description of our model and results.

\subsection{Theoretical challenges related to atomic clocks}

A (passive) atomic clock (See \cite{Riehle, Audoin} for a thorough exposition) consists of a classical local oscillator (LO), e.g. a quartz crystal or a stabilized laser, enslaved to a quantum frequency reference provided by an atomic ensemble. A chosen atomic transition with a frequency $\omega_{ref}$ provides a frequency standard, and the clock time is obtained by the quadrature of the observed local oscillator frequency $\omega_{LO}(t)$,
$$
t_{clock} = \frac{1}{\omega_{ref}} \int_0^t \omega_{LO}(s) \mathrm{d} s.
$$
To obtain a good time keeping device we need to ensure that the relative frequency error $y(t) = (\omega_{LO}(t) - \omega_{ref})/\omega_{ref}$ remains small. To this end an estimation of this error $\hat{y}$ is ascertained at consecutive time intervals of a length $T$, and the local oscillator frequency is then adjusted by $(1- \zeta) \omega_{ref} \hat{y}$, with $(1-\zeta)$ being the gain in the feedback loop. The long time stability of the clock can be captured by the variance
$$
\sigma^2(t):= \frac{1}{t^2} \mathbb{E}[(t- t_{clock})^2], \quad t >>T.
$$

Optical atomic clocks are today's most precise measurement devices with an instability of $10^{-18}$ after $7$ hours of averaging \cite{LudlowR}. Atomic clock accuracy reached a point where the main source of error contributing to $\sigma(t)$ originates in the short time stability of the local oscillator \cite{Ludlow}. This phenomena, of deterioration of the clock time stability due to short time instability of the local oscillator, is often referred to as the Dick effect \cite{Dick, ensembles}. Based on recent experimental data, it was argued that a quantum enhancement of atomic clocks would provide no advantage without improving the stability of the local oscillator \cite{Al-Masoudi}. The main goal of this work is to give a thorough rigorous study of this effect, and to provide quantitative benchmarks for atomic clocks stability with given local oscillator noise.

Theoretical studies of atomic clocks have been concerned with deriving benchmarks for atomic clock stability \cite{Itano, Andre}, devising optimal feedback protocols \cite{Biercuk}, and studying the possibility of enhancing the clocks using entangled states of the frequency reference \cite{Bollinger, Buzek}. There are three obstacles with regard to the latter: decoherence \cite{Huelga}, frequency-phase ambiguity \cite{Rafal} and the local oscillator short time stability. 
To understand  where the above mentioned sources of error originate, we need to take a closer look at the procedure through which the estimation $\hat{y}$ is obtained.

\begin{figure}
\begin{subfigure}{.5\textwidth}
  \centering
  \includegraphics[width=0.9\textwidth]{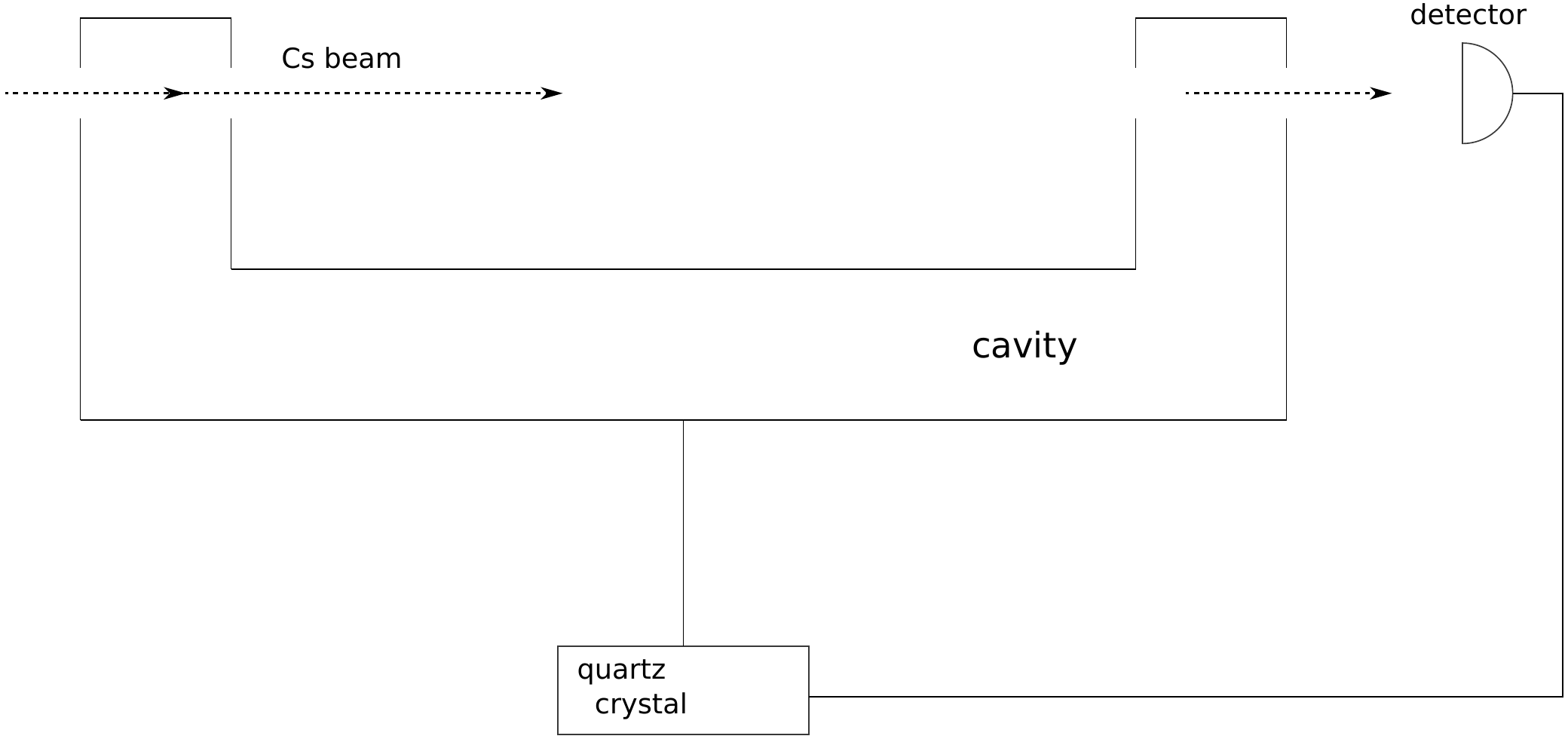}
  \caption{}
  \label{fig:sfig1}
\end{subfigure}%
\begin{subfigure}{.5\textwidth}
  \centering
  \includegraphics[width=0.8\textwidth]{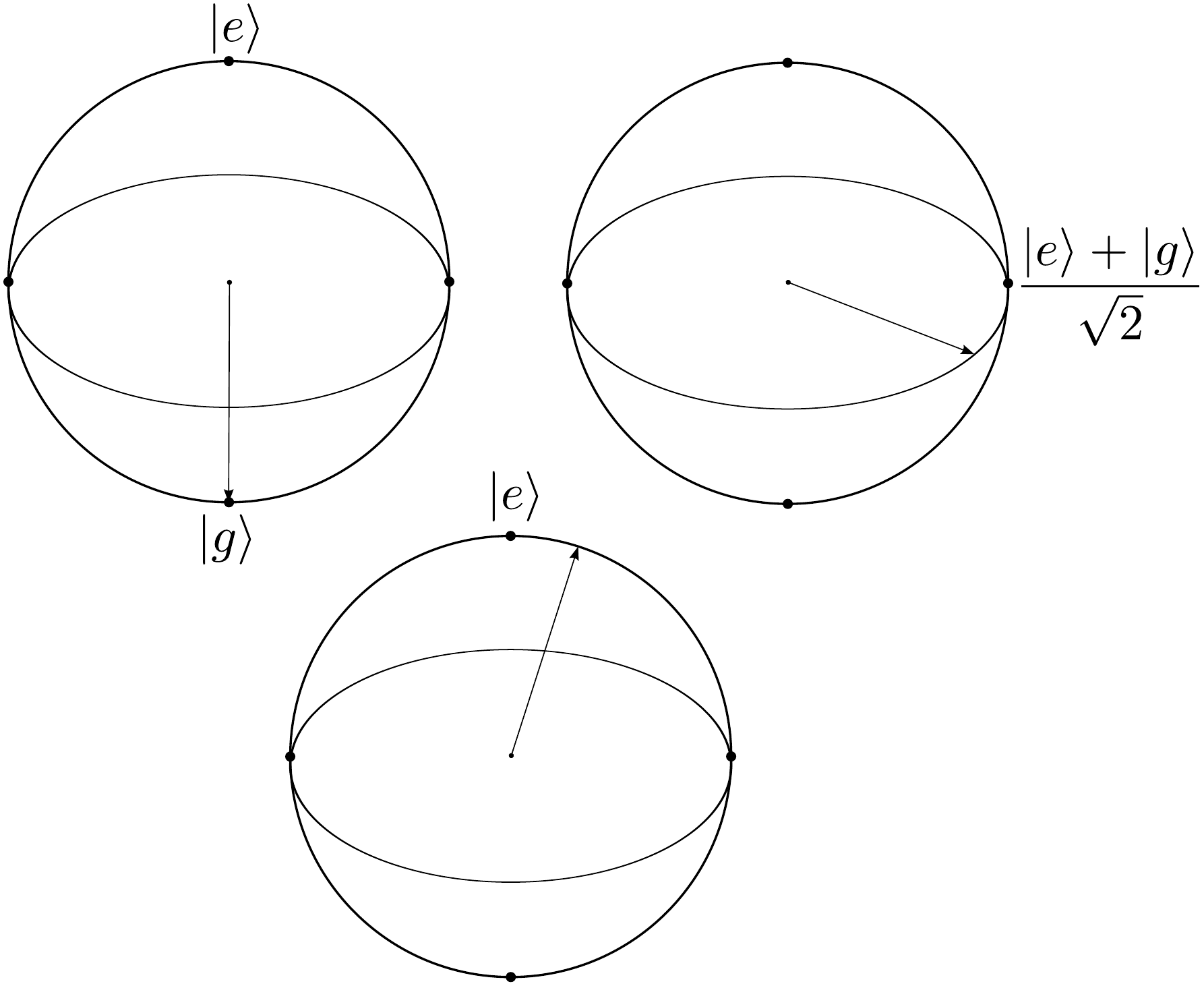}
 \caption{}
  \label{fig:cesium}
\end{subfigure}

\caption{(a) The control scheme of a Cesium atomic clock. A quartz crystal operates an electromagnetic field inside a cavity. A beam of Cesium atoms passes through the cavity into a detector and provides information on the frequency difference between the quartz crystal frequency and the Cesium atom reference frequency. This information is used in a loop to control the quartz crystal in order to make the difference zero. (b) Bloch representation of the state of a Cesium atom during the Ramsey interferometry. Top left: Before entering a cavity the atom is in the ground state. Top Right: In between cavity ends the state is on the equator and undergoes Bloch oscillations with frequency $\omega_{ref}$. Bottom: The state before detection has an angle with the excited state proportional to the acquired relative phase $\int (\omega_{ref} - \omega_{LO}(s))$.}
\label{fig:fig}
\end{figure}

The frequency reference consists of $N$ atoms that are prepared in an initial state $\rho_0$ at the beginning of each interrogation cycle (see Figure~\ref{fig:cesium}). 
During interrogation time, the state rotates with a relative speed proportional to the frequency error $y(s)$ and changes according to
$$
\rho_0 \, \to \, \rho_T(\bar{y}) = e^{-i T \bar{y} H} \rho_0 e^{i T \bar{y} H}, \quad \bar{y} = \frac{1}{T} \int_0^T y(s) \mathrm{d} s,
$$
where $H$ is a Hamiltonian governing the evolution of the atoms. The most common model consists of $N$ non-interacting two level atoms with a Hamiltonian $H = 1/2 \omega_{ref} ( \sigma_{z}^{(1)} + \cdots + \sigma_{z}^{(N)} + N)$, and two paradigmatic initial states $\rho_0 = \ket{\psi} \bra{\psi}$, which are the separable state
\begin{equation}
\label{rev:3}
\ket{\psi_{sep}} = \frac{\ket{0} + \ket{1}}{\sqrt{2}}  \otimes \cdots \otimes \frac{\ket{0} + \ket{1}}{\sqrt{2}},
\end{equation}
and the fully entangled GHZ state
\begin{equation}
\label{rev:2}
\ket{\psi_{GHZ}} = \frac{1}{\sqrt{2}} (\ket{0} \otimes \cdots \otimes \ket{0} + \ket{1} \otimes \cdots \otimes \ket{1}).
\end{equation}
The final state $\rho_T(\bar{y})$ parametrically depends on the relative frequency error, and a measurement on the system provides an estimation $\hat{\bar{y}}$ of this error. The Cramer-Rao bound (\cite{BC} and Section~\ref{QE}) gives a lower bound on the variance of the difference between the estimated and correct values in terms of quantum Fisher information,
$$
\mathbb{E}[(\bar{y} - \hat{\bar{y}})^2] \geq \frac{1}{F}.
$$
In an ideal noiseless situation this bound is asymptotically achievable, and the Fisher information is proportional to the energy variance of the initial state $F = 4 T^2 (\Delta E)^2$. The variance $ (\Delta E)^2 = \bra{\psi} H^2 \ket{\psi} - \bra{\psi} H \ket{\psi}^2$ is equal to $f^2_{ref} N/2$ for the separable state Eq.~(\ref{rev:3}), and it achieves its maximal value $f^2_{ref} (N^2 - N)/2$ for the GHZ state Eq.~(\ref{rev:2}). This is the basic observation behind quantum enhancement estimation schemes, see e.g.\cite{Giovannetti, Rafal2012}.

As mentioned above, in a realistic situation there are three sources of error that severely complicate the noiseless picture. Huelga et. al. \cite{Huelga} pointed out that if the closed system Hamiltonian evolution is replaced by an open system evolution, the $N^2$ asymptotic behavior of the Fisher information for GHZ states changes back to a classical linear asymptotic behavior. This phenomenon has been consequently proved in more general settings (see \cite{RafalK, Escher} and reference therein). The second source of error originates in $\rho_T(\bar{y})$ not depending on the frequency but rather on the phase. In particular, $\rho_T(\bar{y} + \frac{1}{T f_{ref} N^\varepsilon})= \rho_T(\bar{y})$, where $\varepsilon =0$ for the separable state Eq.~(\ref{rev:3}) and $\varepsilon =1$ for the GHZ state. It follows that the GHZ state can distinguish values of $\bar{y}$ only in a region of size $(TN)^{-1}$ around zero. Several recent works studied this problem; Demkowicz-Dobrza{\'n}ski et. al. \cite{Rafal, FMR} found optimal states for a given initial uncertainty of $y$ and developed a framework to derive an optimal interrogation times $T$.  Kessler et. al. \cite{Kessler} implemented a sequential estimation scheme to a model of atomic clocks in which only logarithmic corrections to the optimal Heisenberg scaling are present. The last source of error, and the one most relevant for the present work, stems from the fact that the error provided by estimation, which is only based on information about the average frequency error over the entire interrogation cycle, differs from the actual instantaneous frequency error.

Several recent works studied models that feature some subset of the above errors. Mullan and Knill \cite{Knill, Knill2} devised a dynamical programming algorithm that gives the optimal feedback protocol. Sorensen and Borregaard studied the role of LO noise and entangled states of the frequency reference in the long time stability of the clock time \cite{Sorensen1}, and showed that simultaneous use of multiple frequency references decreases the Dick effect \cite{Sorensen2}. Regarding the latter, the error is not present in the figure of merit given by the relative frequency stability of two clocks \cite{Hume, Takamoto}.

In this work we argue that in order to study the latter two sources of error it is not sufficient to look at a single interrogation cycle. Rather, one needs to determine the stationary process $\omega_{LO}(t)$ when the feedback loop is closed. To this end, and in contrast to the above mentioned works, we study a model with an active feedback.  Even though we focus solely on the third source of error, the ramifications for the phase-frequency ambiguity problem are clear. To determine the probability that the accumulated phase in an interrogation cycle makes a $2 \pi$ flip we need to know the variance of $\omega_{LO}(t)$ in its stationary operation. As we will show below, this variance is sensible to all parameters of the model. 

\subsection{Our model and results}

We proceed to describe a dynamical model that determines the evolution of $y(t)$.  When the feedback loop is open, the evolution of the local oscillator is described by a stochastic flow $K_s y(0)$. We assume that $K_s y(0)$ is a martingale, a choice that encodes the idea that the local oscillator has no knowledge about the reference frequency $\omega_{ref}$. A quantity relevant for the description of this noise is the variance
$$
\sigma^2_{LO}(t) = \mathbb{E}[(\frac{1}{t} \int_0^t K_s y \mathrm{d} s - y)^2].
$$ 
Eventually we will look at an example where $K_s$ is an additive Gaussian noise with a phenomenological ansatz for the variance,
$$
\sigma^2_{LO}(s) = D s^\alpha.
$$
The case $\alpha=-1$ describes a white frequency noise, $\alpha = 0$ the flickr noise, and $\alpha=1$ corresponds to Brownian motion.

When the feedback loop is closed the frequency is adjusted periodically at times $Tn,\,n \in \mathbb{N}$. The relative frequency error $y_n:=y(Tn)$ at the beginning of each interrogation cycle is a Markov process defined by the recursive equation
\begin{equation}
\label{rev:1}
y_{n+1} = K_T y_n - \hat{\bar{y}}_n, \quad \bar{y}_n:=\frac{1}{T} \int_{nT}^{(n+1)T} K_s y_n \mathrm{d} s.
\end{equation}
We study this equation for an arbitrary estimation scheme $\hat{\cdot}$ based on a family of states $\rho_T(y)$ (see Section~\ref{QE} for details). 
 For most of our results we assume that the estimation $\hat{y}$ is proportional to an unbiased estimation, in particular $\mathbb{E}[\hat{y}|y] = (1- \zeta) y$. This is a necessary condition for obtaining an unbiased clock.
 Given a solution $y_n$ of Eq.~(\ref{rev:1}), the associated clock time error is given by
 $$
 t_{clock} - t = T \sum_{n=1}^{t/T} \bar{y}_n.
 $$
 We use  quantum Fisher information,
$$
F(y) := \tr(\rho_T(y) X(y)^2), \quad \frac{1}{2} \{ X(y), \rho_T(y) \} = \frac{\partial \rho_T(y)}{\partial y },
$$
as the information theoretic measure of the frequency reference.

We need to mention two important disambiguations regarding our model. The unbiasedness assumption cannot be satisfied in a model consisting of $N$ atoms because of the frequency-phase ambiguity problem. In fact, it is clear that Eq.~(\ref{rev:1}) cannot have a stationary solution for any non-trivial local oscillator noise and an estimation procedure based on a family of states of $N$ atoms, cf. Eqs.~(\ref{rev:3}, \ref{rev:2}). Indeed, the stationary probability distribution would have to be $\omega_{ref} T N^\varepsilon$  periodic and hence not normalizable (see Section~\ref{sec:out} for an extended discussion). Moreover, Eq.~(\ref{rev:1}) assumes that the local oscillator noise after the feedback is uncorrelated to the local oscillator noise from the previous cycle. This is a Markovian approximation of the feedback model. In a realistic situation an optimal feedback would have to be based on the history of measured data spanning the correlation time of the local oscillator noise \cite{Biercuk}.

A basic result of purely mathematical interest that we prove about Eq.~(\ref{rev:1}) is that the recurrence relation posses a unique stationary solution provided the estimation is unbiased with a gain $1-\zeta$. The main result of our paper (see Eq.~(\ref{CWFRW})) is an inequality for the clock time variance in this stationary state,
\begin{equation}
\label{rev:main}
\lim_{t \to \infty}\frac{\mathbb{E}[(t_{clock} - t)^2]}{t} \geq T \frac{1}{F_T} +  T \sigma^2_{LO}(T) \frac{\beta}{(1-\zeta)^2}.
\end{equation}
The bound on the clock time error consists of two terms. The first term depending on the average Fisher information in the stationary state is universal and describes the quantum projection noise \cite{Itano}. The second depends on the control scheme and the local oscillator noise and corresponded to the Dick effect; $\beta$ is a constant of order one associated with the noise. The inequality can be considered a rigorous justification of the Dick formula. These two terms are also the most relevant for current experiments \cite{Ludlow}.

To obtain a fundamental benchmark solely in terms of the available resources we need to find the optimal interrogation time $T$ that minimizes the RHS of the inequality. On an experimental level, optimizing the interrogation time was suggested to decrease the Dick effect \cite{Westergaard}.  For the phenomenological dependence $\sigma^2_{LO}(T) = D T^\alpha$ with $\alpha > -1$ and the Fisher information associated to a unitary evolution $F_T = 4 T^2 (\Delta E)^2 $, the optimal interrogation time $T$ satisfies the formula
$$
\frac{1}{F_T} \sim \sigma^2_{LO}(T),
$$
which justifies the intuition that the dissipation and the information obtained from the synchronization should be proportional. 
For this optimal time we get an inequality,
$$
\lim_{t \to \infty}\frac{\mathbb{E}[(t_{clock} - t)^2]}{t} \geq \frac{\alpha+2}{\alpha+1} \left( \frac{1}{ 4 \Delta^2 E}\right)^{\frac{\alpha+1}{\alpha+2}} \left(\frac{\beta D (\alpha +1) }{(1-\zeta)^2} \right)^{\frac{1}{\alpha+2}}.
$$
This type of behavior was predicted in \cite[Appendix A]{Wineland}.

Finally, let us demonstrate this inequality by plugging $\Delta^2 E = \omega_{ref}^2 N^{1+ \varepsilon}$, where $\varepsilon =0$ corresponds to a classical scaling, and $\varepsilon =1$ corresponds to the Heisenberg scaling of the Fisher information. The bound then takes the form
$$
\lim_{t \to \infty} \frac{\mathbb{E}[(t_{clock} - t)^2]}{t} \geq Const. N^{-(1+ \varepsilon)\frac{\alpha + 1}{\alpha+2}},
$$
where the constant depends on the reference frequency and noise strength. This in particular shows that the Heisenberg scaling for the clock time stability depends on the local oscillator noise and is given by $N^{-{2} \frac{\alpha +1}{\alpha +2}}$.
The same scaling behavior  was independently derived through a different method by Berry, Hall and Wiseman \cite{Wiseman, Wiseman2} in their study of phase tracking, a problem that is mathematically equivalent to atomic clocks' operation.

In a Gaussian model, see Section~\ref{sec:example}, Eq.~(\ref{rev:1}) takes the form
$$
y_{n+1} = \zeta y_n + (1-\zeta) E_n + K_n,
$$
where $E_n$, resp. $K_n$ are i.i.d. Gaussian random variables with zero mean and variance $F^{-1}_T$, resp. $\beta \sigma^2_{LO}(T)$. These random variables represent the quantum projection noise, resp. the local oscillator noise. This first order autoregressive equation has been studied by Greenhall \cite{Greenhall} as a model of the Dick effect. An elementary computation shows that the equation has a stationary solution with zero mean and a variance
$$
\mathbb{E}[y_n^2] = \frac{1 - \zeta}{1+ \zeta} \frac{1}{F_T} +\sigma^2_{LO}(T) \frac{\zeta^2 + \alpha \zeta + \beta -1 -\alpha}{1 - \zeta^2}.
$$
For the associated clock time an equality holds in the bound (\ref{rev:main}). 

The last part of our work that we want to highlight in the introduction is a conjecture regarding the long time stability of clocks without the unbiasedness assumption. Loosely speaking, the conjecture says that the bound (\ref{rev:main}) holds true provided the local oscillator noise is non-trivial. To substantiate the conjecture we prove the bound without the unbiasedness assumption, while assuming that the process is detailed balance. For the proof of this conjecture we develop a novel global Cramer-Rao inequality, see Eq.~(\ref{DDCR}).

The article is organized as follows. In a preliminary Section~\ref{SP} we recall the basic theory of stochastic processes. In Sections~\ref{ET} and \ref{QE} we briefly describe the classical and quantum estimation theories \cite{Holevo, Helstrom}. In particular, we derive a novel version of the Cramer-Rao bound that emphasizes the role of correlations between an unknown and its estimation. Our model of an atomic clock is fully described in Section~\ref{sec:model}, where we also derive the aforementioned bounds. In Section~\ref{sec:example} we give an example where all bounds are saturated and in Section~\ref{sec:opt} we discuss the optimization of the clock's performance. We close our exposition with outlooks in Section~\ref{sec:out}.

The emphasis in this paper is on studying the aforementioned fundamental properties of our model. This implies, in particular, that we do not aim to prove our statements under minimal conditions. We explicitly assume: 
\begin{assumption}
\label{assumption}
All functions appearing in the text are continuously differentiable in an appropriate space and all probability distributions have a finite second moment.
\end{assumption}

\section{Stochastic processes}
 \label{SP}

We would consider a probability distribution $p(\theta)$ of a single real parameter $\theta$ or a joint probability distribution
$p(\theta,\,\theta')$ of two real parameters $\theta,\,\theta'$. The former is a reduced probability distribution of the latter,  $p(\theta) = \int p(\theta,\,\theta') \mathrm{d} \theta'$.  Furthermore, associated to the latter there is a conditional probability distribution of a single parameter,
$$
p(\theta'|\theta) := \frac{p(\theta, \theta')}{\int p(\theta,\,\theta') \mathrm{d} \theta'},
$$
describing the probability of $\theta'$ given $\theta$. 

For a probability distribution $p(\theta)$ we denote by $\mu(p),\,\sigma(p)$ its mean and variance respectively,
$$
\mu(p):= \int \theta p(\theta) \mathrm{d}\theta, \quad \sigma^2(p):= \int (\theta - \mu)^2 p(\theta) \mathrm{d}\theta.
$$
The mean of a joint probability distribution $p(\theta,\,\theta')$ is the vector of means and its variance is a matrix of mutual covariances. 

Conversely (with a slight abuse of notation), we will often consider pairs of random real-valued variables $\theta,\theta'$ on a probability space\footnote{To simplify the notation we never spell out sigma-algebra explicitly. Those who care should be always able to fill it from the context.} $\{\Omega,\,\mathrm{d}\mu\}$. This induces a join probability distribution $p(\theta,\,\theta')$ that reproduces expectations,
$$
\mathbb{E}[f(\theta,\,\theta')] = \int f(\theta,\,\theta') p(\theta,\,\theta') \mathrm{d}\theta \mathrm{d}\theta'.
$$
The random variable $\theta$ by itself has a probability distribution $p(\theta)$. If random variables are specified only by prescribing their joint probability distribution, then their usage would be independent of a realization (as a function on a certain probability space).

Crucial for estimation theory (and our work) is a notion of conditional expectation. A conditional expectation of $\theta$ given $\theta'$ is a real valued random variable $\mathbb{E}[\theta|\theta']$ on a probability space $\{\Omega,\,\mathrm{d}\mu\}$ given by
$$
\mathbb{E}[\theta|\theta'](x) = \int \theta p(\theta|\theta'(x)) \mathrm{d} \theta, \quad x \in \Omega.
$$
Conditional expectation is a unique random variable measurable with respect to the sigma algebra generated by $\theta'$ (i.e. such that it is constant on the sets where $\theta'$ is constant) that reproduces expectations,
\begin{equation}
\label{ConExp}
\mathbb{E}[f(\theta') \mathbb{E}[\theta|\theta']] = \mathbb{E}[f(\theta') \theta].
\end{equation}
The most useful instance of this formula is $f(x) =1$,
a conditional expectation $\mathbb{E}[\theta|\theta'] $ has the same expectation as $\theta$,  
$$
\mathbb{E}[\theta] = \mathbb{E}[\mathbb{E}[\theta|\theta']].
$$

A space of real valued random variables has a natural associated scalar product $(\theta,\,\theta'):=\mathbb{E}[\theta \theta']$. Random variables of finite variance equipped with this scalar product form a real Hilbert space.  We refer to this scalar product whenever we speak about orthogonality of two random variables.

A stochastic process is a collection of random variables; we will use both stochastic processes, $X_t$, in continuous time $t >0$ and discrete processes $X_n,\,n \in \mathbb{N}$. The first naturally describes frequency dependence on time, the second is a suitable description of measurements occurring in discrete time steps. We would also encounter integrated processes,
$$
\int_0^t X_s \mathrm{d}s,\quad \sum_{j=0}^n X_j.
$$ 
These processes naturally occur as a relation between a clock time and an instantaneous frequency.

Below we consider only discrete processes in details. The corresponding concepts for processes in real time should be clear. 

Of main interest will be the mean and the variance of instantaneous frequency and the variance of the associated clock time. More generally we will frequently use quadratic quantities associated to the process $X_n$. In particular its mean $\mathbb{E}[X_n]$ and autocovariance 
$$
C(X_{n+h},X_n)= \mathbb{E}[(X_{n+h} -\mathbb{E}[X_{n+h}])(X_{n} - \mathbb{E}[x_n])].
$$
For $h=0$ autocovariance reduces to a variance of the process at time $n$.
Quadratic quantities of an integrated process might be computed in terms of integrated covariance. For completeness we give an explicit formula,
%
\begin{multline*}
C(\sum_{j=0}^{n +h} \! X_j,\,\sum_{j=0}^n X_j) = \sum_{j=0}^n C(X_j,X_j) + 2 \sum_{j=0}^n \sum_{k=1}^{n-j} C(X_{j+k},X_j) \\ + \sum_{j=0}^n \sum_{k=n-j+1}^{n+h-j}  C(X_{j+k},X_j).
\end{multline*}

A process $X_n$ is called stationary if a joint distribution of  $X_{n_1 + h}, \cdots X_{n_j +h}$ is independent of $h$. In particular its mean, variance and autocovariance are independent of $n$, we denote $\gamma(h):=C(X_{n+h},X_n)$ and $\gamma(0) = \sigma^2$. The ratio\footnote{In standard notation this would be denoted by $\rho(h)$ however we shall need $\rho$ to denote a quantum state.} $\cor(h):=\gamma(h)/\sigma^2$ is known as a correlation function. The formulas for integrated stationary process simplifies by one summation, e.g.:
$$
C(\sum_{j=0}^{n } \! X_j,\,\sum_{j=0}^n X_j) = (n+1) \sigma^2 + 2 \sum_{h=1}^n  (n-h +1) \gamma(h).
$$
\comment{
\begin{multline*}
C(\int_0^{t +h} \! X_s,\,\int_0^t X_s) = 2 \int_0^t \! \mathrm{d}r\, (t-r) \gamma(r) \\
+\int_t^{t+h} \!\! \mathrm{d}r\, (t+h -r)\gamma(r) + h \int_h^{t} \!\! \mathrm{d}r\, \gamma(r) + \int_0^h \!\! \mathrm{d}r\, r \gamma(r).
\end{multline*}
}
The formula implies that for a stationary process $X_n$ with zero mean, $\mathbb{E}[X_n] = 0$, we have
\begin{equation}
\label{LLN}
\lim_{n \to \infty} \frac{1}{n}\mathbb{E}[(\sum_{j=0}^n X_j)^2] = \sigma^2 + 2 \sum_{h=1}^\infty \gamma(h),
\end{equation}
provided the sum on the RHS converges. In fact, under somewhat more strict conditions on $\gamma(h)$  the central limit theorem gives convergence of $n^{-1/2} \sum_0^n X_j$ to a Gaussian random variable of zero mean and variance given by the RHS of Eq.~(\ref{LLN}).  

Stationary stochastic processes are used for a description of the local oscillator noise. After an initial stage, the frequency of a local oscillator approaches a process that can be described as a mixture of a stationary process and a drift (also called aging). If the latter is negligible, the frequency is a stationary process. Decay of the correlation function $\cor(h)$ is a measure of the stability of the oscillator.

A process $\{X_n\}$ is a martingale if $\mathbb{E}[X_{n+1} | X_n] = X_n$ and it is Markov if the future depends on the past only through the present, $\mathbb{E}[X_{n+1}| X_j,\,j \leq n] = \mathbb{E}[X_{n+1} | X_n]$. The Markov property can be equivalently stated that past and future are independent given the present. This is the first part of following lemma.
\begin{lemma}
\label{XYZ} 
Suppose that $\{X_1,\,X_2,\,X_3\}$ is a Markov chain, then it holds
$$
\mathbb{E}[X_1X_3|X_2] = \mathbb{E}[X_1|X_2] \mathbb{E}[X_3|X_2].
$$
Furthermore when $\mathbb{E}[X_3|X_2] = \zeta X_2$ for some $\zeta \in \mathbb{R}$ then
$$
\mathbb{E}[X_1 X_3] = \zeta \mathbb{E}[X_1 X_2].
$$
\end{lemma}
{\em Proof:}
The first equation is the equivalent definition of Markov property as mentioned in the text above the lemma, see \cite[Chapter II.6]{Doob}. We prove the second part,
\begin{align*}
\mathbb{E}[X_1 X_3] &= \mathbb{E}[\mathbb{E}[X_1 X_3| X_2]] \\
				&= \mathbb{E}[\mathbb{E}[X_1|X_2] \mathbb{E}[X_3|X_2]]\\
				&=\mathbb{E}[\mathbb{E}[X_1|X_2] \zeta X_2] \\
				&=\zeta \mathbb{E}[X_1 X_2].
\end{align*}
In the first and last equality we used Eq.~(\ref{ConExp}). \hfill $\square$

Let $\omega_t$ be a  real valued stochastic process describing a frequency. Then the stability of the frequency source is often described in terms of the standard Allan variance \cite[Chapter 3]{Riehle}
$$
\sigma^2(\tau)= \frac{1}{2 \tau^2} \mathbb{E}[\left( \int_0^\tau y(s) \mathrm{d} s - \int_\tau^{2 \tau} y(s) \mathrm{d} s \right)^2],
$$
where $y(t) = (\omega_t - \mathbb{E}[\omega_t])/\omega_t$ is the relative frequency error and $\tau$ is an averaging time. It is important to note that Allan variance is a function of the averaging time, not a single number. For our purpose Allan variance is unnecessary complicated and throughout the text we will use a simplified quantity that neglects correlations
\begin{equation}
\label{AllanVariance}
\sigma^2(\tau) := \frac{1}{\tau^2} \mathbb{E}[ \left(\int_0^\tau y(s)\mathrm{d}s\right)^2].
\end{equation}
For a stationary process $y(t)$ and $\tau$ large these two quantities coincide.

We end this section with examples of various stochastic processes appearing in the following sections.

\begin{example}[Standard diffusion]
White noise is a stationary process, $X_t$, of uncorrelated random variables. They have a constant mean $\mu$ and autocorrelation function
$$
\gamma(h):=C\left(X_{t+h},X_{t}\right) = D \delta(h).
$$

The integral of white noise, $B_t = \int_0^t X_s$, is a Brownian motion. Its mean and variance are given by formulas
\begin{equation}
\mathbb{E}[B_t] = \mu t, \quad C(B_{t+h}, B_t) = 2 D t. \label{eq:diff}
\end{equation}
A drift $\mu$ and a diffusion coefficient $D$ are constants whose physical dimension ($[\,\cdot\,]$) depends on the process. More precisely,
$[\mu] = [X_s]$, $[D] = [X_s]^2$.

Brownian motion $B_t$ is a continuous martingale.
\end{example}

\begin{example}[Gaussian random process]
\label{GSP}
A discrete process $X_n$ is called Gaussian if the joint probability distribution of $X_{n_1},\,X_{n_2},\dots,X_{n_j}$ is a multivariate normal distribution for any $j$-tuple $n_1,\,\dots n_j$.

The Gaussian process is completely determined by the means $\mu(X_n)$ and covariances $C(X_n,\,X_m)$. A particular property of interest (see \cite[Chapter 7.3]{Cox}) is that for a stationary Gaussian processes with zero mean and variance $\sigma^2$ it holds that
$$
\mathbb{E}[X_{n+h} X_n] = \zeta^h \sigma^2,
$$
where $\zeta =\mathbb{E}[X_{n+1} X_n]/\sigma^2$.  
\end{example}

\begin{example}[Exponentially decaying correlations]
\label{exa:DOC}
A stationary discrete stochastic process has exponentially decaying correlations if for some $|\zeta|<1$,
$$
\mathbb{E}[X_{n+h} X_n] = \zeta^h \sigma^2.
$$
The variance of the associated integrated process can be compute explicitly by summing a geometric series. Note that the result is consistent with Eq.~(\ref{LLN}).
\begin{align}
\label{EDOCA}
\mathbb{E}[\left( \sum_{n=0}^N X_n \right)^2]    &= (N+1) \sigma^2 + 2 N \sigma^2 \frac{\zeta}{1 -\zeta } \left( 1 + \frac{1}{N} \frac{\zeta(\zeta^{N}-1)}{1- \zeta}\right) \nonumber \\ 
									&= N\sigma^2 \frac{1+\zeta}{1 - \zeta}  + O(1).
\end{align}
\end{example}

\section{Estimation theory}
\label{ET}
Estimation theory studies strategies how to estimate an unknown physical parameter $\varphi$ based on a data collected from a single or multiple measurements. In the classical estimation theory there is usually a one to one correspondence between an ideal measurement and the unknown. The problem is then to decrease a measurement error using large data sets.
Our exposition of the estimation theory would be directed towards application in atomic clocks. Reader can find a general reference in e.g. \cite{Deutsch}, \cite{Berger}.

We examine strategies to estimate a parameter $\varphi \in \mathbb{R}$ based on a measurement outcome $\mu$. The space of outcomes, $\mathcal M$, is a probability space equipped with a measure $\mathrm{d} \mu$, and a probability distribution $p(\mu| \varphi)$ encodes the probability of an outcome $\mu$ for a given $\varphi$.
 The estimation strategy is then defined by  an estimator $\Phi$. Upon a measurement outcome $\mu$ a guess $\Phi(\mu)$ is made. $\Phi$ is a function from the space of outcomes to real numbers.

In a Bayesian approach to the estimation theory $\varphi$ is a random variable, on a probability space $(\Omega, \mathrm{d} \varphi)$, with a certain prior probability distribution $q(\varphi)$. It is then convenient to view $\mu$ as a random variable on the joint probability space $(\Omega \oplus \mathcal{M}, p(\mu | \varphi) \mathrm{d} \varphi \mathrm{d} \mu)$ given by the coordinate projection on the space $\mathcal{M}$.
 
 \begin{definition}[Estimation]
 \label{def:est}
 Let $\varphi$ and $\mu$ be random variables defined in the above paragraph and $\Phi : \mathcal{M} \to \mathbb{R}$ an estimator. Then an estimation $\hat \varphi$ of $\varphi$ is a real valued random variable
 $$
 \hat \varphi := \Phi \circ \mu .
 $$
 
 In explicit terms, this is a random variable
 $$
 \hat \varphi\, :\,\left(\Omega \oplus \mathcal{M}, p(\mu | \varphi) \mathrm{d} \varphi \mathrm{d} \mu\right) \rightarrow \mathbb{R}
 $$
 given by
 $$
 \hat \varphi(\varphi,\mu) = \Phi(\mu).
 $$
 \end{definition}
 
It is common to denote the estimator $\Phi$ and the estimation $\hat \varphi$ by the same letter. This is indeed convenient if $\varphi$ is fixed with a given prior distribution. However we will consider estimations of a chain of random variables based on a fixed estimator $\Phi$. For that reason we prefer to stress in our notation that $\hat \varphi$ depends on the random variable that is estimated while $\Phi$ is a fixed function. 
 
 Unbiased estimators play a central role in the estimation theory. In the rest of this subsection, which is devoted to their exposition, we fix a conditional probability distribution $p(\mu|\varphi)$.
  \begin{definition}[Unbiased estimation]
  We say that an estimator $\Phi$ is $\zeta$-biased if for all random variables $\varphi$
  $$
  \mathbb{E}[\hat \varphi | \varphi] = (1-\zeta) \varphi.
  $$
  The estimator is unbiased if it is $0$-biased.
  We also say that an estimator is conditionally unbiased if 
  $$
  \mathbb{E}[\varphi] = 0 \quad \implies \quad \mathbb{E}[\hat \varphi] = 0.
  $$
  \end{definition}
  
A $\zeta$-biased estimation (for $\zeta \neq 0$) is not a common concept, in fact a $\zeta$-biased estimator is proportional to an unbiased estimator; however the parameter $\zeta$ will correspond to the gain in a feedback loop and hence play an important role in our description of an atomic clock. Note also that $\{\varphi,\hat\varphi + \zeta \varphi \}$ is a martingale, $\mathbb{E}[\hat \varphi + \zeta \varphi | \varphi] = \varphi$. In particular we will often use that for $\zeta$-biased estimation
\begin{equation}
\label{orth}
\mathbb{E}[\varphi (\hat \varphi - (1-\zeta) \varphi)] = 0.
\end{equation} 
The equation follows from the stated conditional expectations and Eq.~(\ref{ConExp}).

The $\zeta$-biased property of estimation can be equivalently stated by referring only to the conditional probability distribution $p(\hat \varphi| \varphi)$. Consequently we often say that $\hat \varphi$ is a $\zeta$-biased estimation of $\varphi$, meaning that it is an estimation of the unknown $\varphi$ based on a $\zeta$-biased estimator.

 The following lemma summarizes various useful statements about unbiased estimators.
\begin{lemma}
\label{lem:COU}
For an estimator $\Phi$ the following is equivalent
\begin{enumerate}
\item[(i)] $\Phi$ is conditionally unbiased,
\item[(ii)]there exists $\zeta \in \mathbb{R}$ such that $\Phi$ is $\zeta$-biased estimator.
\end{enumerate}
Suppose in addition that $\zeta \neq 1$. Then a $\zeta$-biased estimator $\Phi$ has the form $\Phi = (1-\zeta) \Phi_0$ where $\Phi_0$ is an unbiased estimator.
\end{lemma}
 {\em Proof:}
 (i) $\implies$ (ii): Let $p(\hat \varphi | \varphi)$ be a conditional probability distribution of $\hat \varphi$ given $\varphi$ and let $q(\varphi)$ be a probability distribution of $\varphi$. Then (i) states that for all distributions $q(\varphi)$ with zero mean it holds
 $$
 \int \hat \varphi p(\hat \varphi | \varphi) q(\varphi) \mathrm{d} \hat \varphi \mathrm{d} \varphi = 0.
 $$
 A standard variational argument implies 
 $$
 \int \hat\varphi p(\hat \varphi | \varphi) \mathrm{d}\hat \varphi = \zeta \varphi
 $$
 for some $\zeta \in \mathbb{R}$. This is exactly (ii).
 
 (ii) $\implies$ (i): For a random variable $\varphi$ with zero mean and $\zeta$-biased estimation $\hat \varphi$ it holds
 $$
 \mathbb{E}[\hat \varphi] = \mathbb{E}[\mathbb{E}[\hat \varphi|\varphi]]  = (1- \zeta) \mathbb{E}[\varphi] = 0.
 $$ 
 
 When $\Phi$ is a $\zeta$-biased estimator and $\zeta \neq 1$ then $(1 -\zeta)^{-1} \Phi$ is clearly an unbiased estimator.
 \hfill $\square$.
  
  Most of the work in estimation theory is centered on minimizing certain cost of $\varphi - \hat\varphi$ not hitting zero. We discuss this in the following section.
  
\subsection{A cost of the estimation}
  \label{sec:Cost}
 A cost of the estimation (i.e. a functional we aim to minimize) is given by
 \begin{align}
 Cost &= \mathbb{E}[(\varphi - \hat\varphi)^2] \nonumber \\
          &= \int (\varphi - \Phi(\mu))^2 p(\mu|\varphi) q(\varphi) \mathrm{d} \mu \mathrm{d} \varphi \label{Cost},
 \end{align}
 where $q(\varphi)$ is a prior probability distribution of $\varphi$.
The choice of the cost function is to a large extent arbitrary. The quadratic cost function is distinguished by its simplicity and a direct relation to variance, the quantity that is most suitable for a description of the time precision.

It is well known how to optimize the cost, Eq.~(\ref{Cost}), with respect to the estimator $\Phi$ for a fixed prior distribution of the variable $\varphi$.
\begin{lemma}[Optimal estimator]
\label{lem:OptEst}
Fix a conditional probability distribution $p(\mu|\varphi)$ and a prior distribution $q(\varphi)$.  Then an estimator
\begin{align}
\label{OptEst}
	\Phi(\mu) &= \mathbb{E}[\varphi | \mu] \\
			       &=\int \varphi p(\varphi|\mu) \mathrm{d} \varphi \nonumber
\end{align}
minimizes the cost (\ref{Cost}) with respect to the estimator $\Phi(\cdot)$.
\end{lemma}

{\em Proof:} We use the formula $\mathbb{E}[Z(\mu) Y] = \mathbb{E}[Z(\mu) \mathbb{E}[Y|\mu]]$ twice to rewrite the cost as
\begin{align*}
\mathbb{E}[(\Phi(\mu) - \varphi)^2] &= \mathbb{E}[\Phi(\mu)^2 - 2\Phi(\mu) \varphi + \varphi^2] \\
								&=\mathbb{E}[\Phi(\mu)^2 - 2\Phi(\mu) \mathbb{E}[\varphi|\mu] + \varphi^2] \\
								&=\mathbb{E}[(\Phi(\mu) - \mathbb{E}[\varphi|\mu])^2] + \mathbb{E}[(\mathbb{E}[\varphi|\mu] - \varphi)^2].
\end{align*}
The last expression is a sum of two squares, where the second is independent of $\Phi$. Hence the minimum is achieved when the first square vanishes. \hfill $\square$

The explicit expression, Eq.~(\ref{OptEst}) is often hard to analyze. This is the case when the conditional probability distribution $p(\mu|\varphi)$ has an analytical expression, however there is no such expression for the conditional probability distribution $p(\varphi|\mu)$. In such cases bounds of the cost from below are very useful. Of such bounds the most famous is the Cramer-Rao bound, a variant of which we present here. It bounds the cost from below in terms of the Fisher information. This is a point-wise quantity that (roughly speaking) measures how fast does a conditional probability distribution changes with the value of the condition. 

The Fisher information, $F(\varphi)$, associated to a probability distribution $p(\mu|\varphi)$ is given by
\begin{equation}
\label{Fclas}
F(\varphi) := \int \left( \frac{\partial}{\partial \varphi} \log p(\mu|\varphi) \right)^2 p(\mu|\varphi) \mathrm{d} \mu.
\end{equation}
An important property of the Fisher information is that it decreases by processing of the information. For any fix estimator $\Phi$, the Fisher information associated to the conditional probability distribution $p(\hat \varphi| \varphi)$ is always less than equal to the Fisher information associated to the family $p(\mu | \varphi)$, 
\begin{equation}
\label{fphi}
F(\varphi) \geq \int \left( \frac{\partial}{\partial \varphi} \log p(\hat \varphi|\varphi) \right)^2 p(\hat \varphi|\varphi) \mathrm{d} \hat \varphi.
\end{equation}
We will use this inequality repeatedly in the proofs of this section without a further comment.

The original Cramer-Rao bound (that we present in an integrated version) is the following statement.
\begin{proposition}
Suppose that $\hat \varphi$ is an unbiased estimation (i.e. an estimation based on an unbiased estimator) of a random variable $\varphi$. Then
\begin{equation}
\label{CCR}
\mathbb{E}[(\varphi - \hat\varphi)^2] \geq \mathbb{E}[\frac{1}{F(\varphi)}].
\end{equation}
\end{proposition}
{\em Proof:}
For an unbiased estimation a conditional probability, $p(\hat\varphi | \varphi)$, of $\hat \varphi$ given $\varphi$ satisfies
$$
\int \hat\varphi p(\hat\varphi|\varphi) \mathrm{d} \hat\varphi = \varphi.
$$
We differentiate the expression, subtract zero and use the Cauchy-Schwarz inequality
\begin{align}
1 &= \left(\int (\hat\varphi - \varphi) \partial_\varphi p(\hat\varphi|\varphi) \mathrm{d} \hat\varphi \right)^2 \nonumber \\
   &\leq F(\varphi) \int (\hat \varphi - \varphi)^2 p(\hat \varphi | \varphi) \mathrm{d}\hat\varphi . \label{eq:CS}
\end{align}
Dividing by $F(\varphi)$ gives a pointwise version of the inequality, Eq.~(\ref{CCR}) can be then obtained by applying $\mathbb{E}[\cdot]$ to both sides. \hfill $\square$

An immediate corollary is a bound for $\zeta$-biased estimation.
\begin{corollary}
Suppose that $\hat \varphi$ is a $\zeta$-biased estimation of a random variable $\varphi$. Then
\begin{equation}
\label{CRUZ}
\mathbb{E}[(\varphi - \hat\varphi)^2] \geq (1-\zeta)^2 \mathbb{E}[\frac{1}{ F(\varphi)}] + \zeta^2 \mathbb{E}[\varphi^2].
\end{equation}
\end{corollary}
{\em Proof:}
For $\zeta \neq 1$ an estimation $\hat \varphi/(1-\zeta)$ is unbiased and the statement follows from
\begin{align*}
\mathbb{E}[(\varphi - \hat\varphi)^2] &= (1-\zeta)^2 \mathbb{E}[(\varphi - \frac{\hat \varphi}{1-\zeta})^2] + \zeta^2 \mathbb{E}[\varphi^2] \\
							&\geq (1-\zeta)^2 \mathbb{E}[\frac{1}{ F(\varphi)}] + \zeta^2 \mathbb{E}[\varphi^2] ,
\end{align*}
where the equality in the first line follows from orthogonality of $(1-\zeta)\varphi - \hat\varphi$ and $\varphi$, see Eq.~(\ref{orth}).

The case $\zeta =1$ is somewhat special\footnote{And completely unimportant.}. In view of $\mathbb{E}[\varphi \hat \varphi] = 0$ it then holds
$$
\mathbb{E}[(\varphi - \hat\varphi)^2] = \mathbb{E}[\varphi^2] + \mathbb{E}[\hat\varphi^2]
$$
and we see that the optimal estimation is $\hat\varphi = 0$.
\hfill $\square$

Van Trees \cite{VanTrees} proved a Cramer-Rao type bound for an arbitrary estimator. We give a version of this bound,  Eq.~(\ref{DDCR}), that generalizes Eq.~(\ref{CRUZ}) and which to the best of our knowledge is new. It recognizes a role of correlations between $\varphi$ and $\hat \varphi$ in the Cramer-Rao inequality.

An extension of Cramer-Rao inequality beyond unbiased estimators comes at the expense of a less natural averaging of the Fisher information or introduction of additional terms. We choose the former approach because it has the simplest proof and gives the nicest formulas; A note on the other approach would be given elsewhere.
For a given probability distribution $q(\varphi)$ we introduce an average Fisher information
\begin{equation}
\label{AF}
\tilde F = \int F(\varphi) \frac{\tilde{q}(\varphi)^2}{q(\varphi)} \mathrm{d}\varphi, 
\end{equation}
where
\begin{equation}
\label{temp78}
\tilde{q}(\varphi)= \frac{\int_{\varphi}^\infty (s - \mu(q)) q(s) \mathrm{d}s}{\sigma(q)^2}
\end{equation}
is  a probability distribution associated to $q(\varphi)$.

For simplicity of the exposition we assume in the following theorem that $\varphi$ has zero mean. This is also the only case we will use in the article.
\begin{thm}
\label{thm:DDCR}
Let $\hat \varphi$ be an estimation of a random variable $\varphi$ (of zero mean) with a prior probability distribution $q(\varphi)$. Denote $\zeta := \mathbb{E}[(\varphi - \hat \varphi)\varphi] /  \mathbb{E}[\varphi^2]$. Then it holds
\begin{equation}
\label{DDCR}
\mathbb{E}[(\varphi - \hat\varphi)^2] \geq (1 - \zeta)^2 \frac{1}{\tilde F} + \zeta^2 \mathbb{E}[\varphi^2],
\end{equation}
where $\tilde F$ is the average Fisher information, Eq.~(\ref{AF}).
\end{thm} 

{\em Proof:}
The definition of $\zeta$ implies that variables $(1-\zeta)\varphi - \hat\varphi$ and $\varphi$ are orthogonal with respect to a natural scalar product.  This suggest (and proves) a decomposition
\begin{equation}
\label{decomposition}
\mathbb{E}[(\varphi - \hat\varphi)^2] = \mathbb{E}[(\hat\varphi - (1-\zeta) \varphi)^2] + \zeta^2 \mathbb{E}[\varphi^2].
\end{equation}
Now we bound the first term on the RHS. Using the definition of $\zeta$ once again we have
$$
\int \hat \varphi p(\hat \varphi | \varphi) \varphi q(\varphi) \mathrm{d}\varphi \mathrm{d}\hat\varphi = (1-\zeta) \mathbb{E}[\varphi^2].
$$
It follows by integration by parts that for any $a \in \mathbb{R}$ (the term proportional to $a$ is point-wise zero)
$$
\int(\hat \varphi -a \varphi) \frac{\partial}{\partial \varphi} p(\hat \varphi|\varphi) \int_\varphi^\infty s q(s) \mathrm{d}s \,\mathrm{d}\varphi \mathrm{d}\hat \varphi = (1-\zeta) \mathbb{E}[\varphi^2].
$$
This further implies (note a definition of $\tilde{q}$, Eq.~(\ref{temp78}))
\begin{align*}
(1-\zeta)^2 &= \left(\int(\hat \varphi -a \varphi) \frac{\partial}{\partial \varphi} p(\hat \varphi|\varphi)\tilde{q}(\varphi)\mathrm{d}\varphi \mathrm{d}\hat\varphi \right)^2 \\
		&\leq \mathbb{E}[(\hat\varphi -a \varphi)^2] \int F(\varphi) \frac{\tilde{q}(\varphi)^2}{q(\varphi)} \mathrm{d} \varphi \\
		&\leq \mathbb{E}[(\hat\varphi -a \varphi)^2] \tilde F.
\end{align*}
Inserting this into Eq.~({\ref{decomposition}) proves the sought inequality. \hfill $\square$

The inequality, Eq.~({\ref{DDCR}), naturally bridges between the classical Cramer-Rao inequality for an unbiased estimator and a global Cramer-Rao inequality. 
To see this note that minimizing over $\zeta$ on the right hand side gives us a clone of Van Trees inequality (see \cite{Gill}),
\begin{align*}
\mathbb{E}[(\varphi - \hat\varphi)^2] 		&\geq \inf_{\zeta} \left( (1 - \zeta)^2 \frac{1}{\tilde F} + \zeta^2 \mathbb{E}[\varphi^2]\right) \\
								&= \frac{1}{\tilde F + 1/\mathbb{E}[\varphi^2] }.
\end{align*}
On the other hand $\zeta = 0$ reproduces Eq.~(\ref{CCR}) up to a different averaging of the Fisher information.

The special averaging of Theorem~\ref{thm:DDCR} is very suitable for Gaussian prior distributions.

\begin{example}
Suppose that  the prior distribution $q(\varphi)$ of a random variable $\varphi$ is Gaussian,
then $\tilde{F}$ is an average Fisher information with respect to the distribution $q$, i.e.
$$
\tilde{F} = \mathbb{E}[F(\varphi)].
$$
Note however that by the Jensen inequality 
$$
\mathbb{E}[\frac{1}{F(\varphi)}] \geq \frac{1}{\mathbb{E}[F(\varphi)]}
$$
and so if an estimator is $\zeta$-biased the Cramer-Rao inequality (\ref{CRUZ})  gives better bound than (\ref{DDCR}) even in this case.  Inequalities coincide only if we further assume that the Fisher information $F(\varphi)$ is constant. 
\end{example}

{\em Proof:} One can directly verify that $\tilde{q}(\varphi)$ of Eq.~(\ref{temp78}) associated to a Gaussian distribution $q(\varphi)$ satisfies $\tilde{q}(\varphi) = q(\varphi)$. \hfill $\square$

\section{Quantum estimation theory}
\label{QE}
 In contrast to the classical estimation theory, quantum measurements cannot distinguish between non-orthogonal states even in the ideal situation of no external noise. This gives a fundamental bound on estimation precision which is referred to as the Heisenberg limit. Unlike the classical case, the probability distribution of this intrinsic quantum measurement error is described by the theory itself.  
 
 Throughout the text we fix a Hilbert space $\mathcal{H}$ representing the quantum system. A state $\rho$ on $\mathcal{H}$ is a positive operator of unit trace. A pure state is represented by a one dimensional projection, which we mostly denote by $P$. A POVM measurement is defined by operators $\Pi(\mu) \geq 0$ that decompose the identity, $\int \Pi(\mu) \mathrm{d}\mu = 1$. The probability distribution (with respect to a measure $\mathrm{d} \mu$) of a measurement outcome $\mu$ given the state $\rho$ is given by the standard formula $\tr(\rho \Pi(\mu))$. 
  
 We examine strategies to estimate a parameter $\varphi \in \mathbb{R}$ of a quantum state $\rho(\varphi)$, whose dependence on the parameter $\varphi$ is known. The estimation strategy is defined by a POVM measurement $\Pi(\mu)$ and an estimator $\Phi(\mu)$. A POVM measurement $\Pi(\mu)$ induces a conditional probability distribution of measurement outcomes $p(\mu|\varphi) = \tr(\Pi(\mu) \rho(\varphi))$ and hence for every fixed POVM measurement we obtain a well posed classical estimation problem. Consequently for each fixed $\{\Pi(\mu), \Phi\}$ the estimation $\hat{\varphi}$ of $\varphi$ is defined (see Definition~\ref{def:est}), and we say that it is $\zeta$-biased if for all random variables $\varphi$ it holds that $\mathbb{E}[\hat \varphi| \varphi] = (1-\zeta) \varphi$.
 
Let $\{\Pi(\mu), \hat \varphi\}$ be an estimation strategy. Then a conditional probability distribution function $p(\hat\varphi | \varphi)$ of $\hat\varphi$ conditioned upon $\varphi$ is given by (we assume that $|\grad \Phi| > 0$ and use a coarea formula)
 \begin{equation}
 \label{coarea}
 p(\hat \varphi | \varphi) = \int_{\Phi^{-1}(\hat \varphi)} \tr\left(\rho(\varphi) \Pi(x)\right) |\grad \Phi(x)|^{-1} \mathrm{d} \nu(x),
 \end{equation}
 where $\mathrm{d} \nu$ is the induced measure by $\mathrm{d} \mu$ on the manifold $\Phi^{-1}(\hat \varphi)$. In particular we see that a POVM measurement with outcomes $\hat \varphi \in \mathbb{R}$ given by 
 $$\tilde{\Pi}(\hat \varphi) =\int_{\Phi^{-1}(\hat \varphi)}  \Pi(x) |\grad \Phi(x)|^{-1} \mathrm{d} \nu(x)$$ 
 and an identity estimator function is equivalent to the original pair $\{\Pi(\mu),\,\Phi\}$.
 The equivalence of these two pairs can be also explained in a down to earth language:
 The label $\mu$ of the measurement outcome is a superficial quantity and we can always re-parameterize it so that the measurement outcome is the estimation itself. In particular $\tr(\tilde\Pi(\hat \varphi) \rho(\varphi))$ is a conditional probability of $\hat \varphi$ given $\varphi$. 
 
In view of the previous paragraph we will assume throughout the text that $\Phi = 1$ and $\mu \equiv \hat{\varphi} \in \mathbb{R}$. The estimation strategy is then defined by the POVM $\Pi(\hat\varphi)$, which incorporates the estimator function $\Phi$.  
The cost of the estimation is now given by, cf. Eq.~(\ref{Cost}),
\begin{align*}
Cost &= \mathbb{E}[(\varphi - \hat\varphi)^2] \nonumber \\
          &= \int (\varphi - \hat\varphi)^2 \tr\left(\Pi(\hat\varphi) \rho(\varphi)\right) q(\varphi) \mathrm{d} \hat\varphi \mathrm{d} \varphi ,
\end{align*}
  where $q(\varphi)$ is a prior distribution of a random variable $\varphi$. In the rest of this section we fix the family $\rho(\varphi)$ and discuss the dependence of the cost on the estimation scheme $\Pi(\hat\varphi)$.
  
In the classical case, Lemma~\ref{lem:OptEst} describes optimization of the cost with respect to the estimation scheme. A quantum equivalent of this Lemma have been derived in \cite{Helstrom, Rafal}, the optimal POVM is  orthogonal and there exists a closed form algebraic expression for an observable associated to this measurement. This expression is, however, hard to analyze and we resort to a more tractable expressions that bound the cost from below.

The quantum Cramer-Rao bound is a generalization of the classical one. It bounds the cost from below using the (quantum) Fisher information, which is a canonical statistical length on the space of density matrices.  
The Fisher information $F(\varphi)$ is given by the expression
$$
F(\varphi) = \tr(\rho(\varphi) X(\varphi)^2),
$$
where $X(\varphi)$ is a solution\footnote{The equation does not determine $Q X(\varphi) Q$, where $Q$ is the orthogonal projection on $\mathrm{Ker}(\rho(\varphi))$, and this part of $X(\varphi)$ can be chosen arbitrary.}  of an equation
$$
\frac{1}{2}\{X(\varphi),\,\rho(\varphi)\} = \dot{\rho}(\varphi), \qquad (\cdot = \frac{d}{d \varphi}),
$$
the expression $\{A,\,B\} = AB + BA$  is the anti-commutator of operators $A$ and $B$.
When $\rho(\varphi) \equiv P(\varphi)$ is a family of projections then $X(\varphi) = \dot{P}(\varphi)$ and Fisher information is proportional to the Fubini-Study metric, $F(\varphi) = 2 \tr(\dot{P}(\varphi)^2)$. 

Braunstein and Caves \cite{BC} give a connection between the classical Fisher information associated to a fix measurement $\Pi(\hat \varphi)$ and the quantum Fisher information. 
\begin{proposition}
Consider a family of states $\rho(\varphi)$ and POVM measurements $\Pi(\hat\varphi)$. Let $F(\varphi)$ be the quantum Fisher information associated to the family $\rho(\varphi)$ and let $F_\Pi(\varphi)$ be the Fisher information, Eq.~(\ref{Fclas}), associated to the conditional probability distribution
$$
p(\hat\varphi|\varphi) = \tr(\Pi(\hat\varphi) \rho(\varphi)).
$$
Then it holds that
$$
F(\varphi) = \sup_{\Pi} F_\Pi(\varphi),
$$
where the supremum is taken over all POVM measurements $\Pi(\hat\varphi)$.
\end{proposition}
{\em Proof:} Let $X$ be a hermitian operator and $A,\,B$ non-negative operators, then a Cauchy-Schwartz inequality
$$
|\tr(XAB)| = |\tr(B^\frac{1}{2} X A^\frac{1}{2} A^\frac{1}{2} B^\frac{1}{2})| \leq \sqrt{\tr(AB) \tr(BXAX)}
$$
combined with the same inequality for $A$ and $B$ exchanged imply
$$
\left| \tr \left(X \{A,B\} \right) \right|^2 \leq 4 \tr (AB) \tr(AXBX).
$$

We use the latter inequality for $X = X(\varphi),\,A=\rho(\varphi)$ and $B= \Pi(\hat\varphi)$, (we omit the arguments of the operators)
\begin{align*}
|\tr(\Pi \dot \rho)|^2 = |\tr(\Pi \{X, \rho\})|^2 &= |\tr(X \{ \Pi, \rho \} )|^2 \\
						& \leq 4 \tr(\Pi \rho) \tr(X \rho X \Pi) .
\end{align*}
Hence we have the following estimate for the classical Fisher information,
\begin{align*}
F_{\Pi}(\varphi) &= \int \frac{(\tr( \Pi(\hat\varphi) \dot \rho(\varphi))^2}{\tr(\Pi(\hat\varphi) \rho(\varphi))} \mathrm{d} \hat\varphi \\
			&\leq \int 4 \tr(X(\varphi)\rho(\varphi) X(\varphi) \Pi(\hat\varphi)) \mathrm{d} \hat\varphi = F(\varphi),
\end{align*}
the last expression being the quantum Fisher information. Equality can be achieved by taking $\Pi(\hat\varphi)$ as a spectral decomposition of $X(\varphi)$. \hfill $\square$.

All classical versions of the Cramer-Rao bound then immediately imply their quantum counterparts. We present one as an example, which is a compilation of bounds (\ref{CRUZ}) and (\ref{DDCR}). In parallel to the classical case we define $\tilde{F}$ with respect to a probability distribution $q$ by Eq.~(\ref{AF}).
\begin{thm}
Consider a family of states $\rho(\varphi)$ and let $F(\varphi)$ be the associated quantum Fisher information. Let $\hat \varphi$ be an estimation of a random variable $\varphi$ (of zero mean) with a prior distribution $q(\varphi)$ and denote $\zeta:=\mathbb{E}[(\varphi - \hat\varphi) \varphi] / \mathbb{E}[\varphi^2]$. Then it holds
\begin{equation}
\label{QDDCR}
\mathbb{E}[(\varphi - \hat \varphi)^2] \geq (1-\zeta)^2 \frac{1}{\tilde{F}} + \zeta^2 \mathbb{E}[\varphi^2],
\end{equation}
where $\tilde F$ is an average Fisher information, Eq.~(\ref{AF}).

If furthermore $\hat \varphi$ is an unbiased estimation then the term $1/\tilde{F}$ in the inequality can be replaced by a simple average $\mathbb{E}[1/F(\varphi)]$.
\end{thm}

\begin{example}[Hamiltonian family]
\label{HEvolution}
Let $P(\varphi) = e^{-i \varphi H} P e^{i \varphi H}$ be a family of pure states generated by a Hamiltonian $H$. Then the Fisher information is constant and proportional to the variance of the energy,
$$
F(\varphi) = 4 \left( \tr(H^2 P) - \tr^2(H P) \right).
$$
The Cramer-Rao inequality then takes a form of the Heisenberg uncertainty relation.
\end{example}

\section{A model of atomic clocks}
\label{sec:model}
In our model, the clock is fully described by a relative frequency error $y(t)$. It is a real valued random variable and the purpose of this section is to define the process $\{y(t)\}_{t \geq 0}$ and to discuss its basic properties. Clock's frequency and the  clock time are then determined through
$$
y(t) = \frac{\omega_{LO}(t) - \omega_{ref}}{\omega_{ref}}, \qquad t_{clock} - t = \int_0^t y(s) \mathrm{d}s.
$$ 

The model consists of various parameters/objects that determine the process; we list them here:
\begin{itemize}
\item Time between two consecutive synchronizations, $T$.

\item A Markovian stochastic process $K_t  \varphi$ that describes evolution of the error in absence of synchronization given an initial condition $K_0 \varphi = \varphi$. 

\item A family of states $\rho_T(\varphi)$ and an estimation strategy $\Pi(\hat\varphi)$. The family describes the state of the frequency reference after the interrogation and its dependence on $T$ and $\varphi$ is prescribed. In concrete examples, the dependence is determined by a closed, or open, system evolution of the state of the  reference. 

\end{itemize}

The adjustment of error after the synchronization is performed periodically at times $n T,\, n\in \mathbb{N}$. We denote
\begin{align}
y_n(t)&:=y(n T + t), \quad \mbox{for} \quad t \in [0,\,T), \nonumber\\
\bar y_n &:= \frac{1}{T} \int_{0}^T y_n(s) \mathrm{d} s. \label{taua}
\end{align}
We also abbreviate $y_n:=y_n(0) = y(nT)$. The stochastic process $y(t)$ is defined implicitly by an initial condition $y_0$ through equations
\begin{align}
	y_n(t) &= K_t y_n \quad \mbox{for} \quad t \in [0,T), \label{EAbs} \\
	y_{n+1} &= K_T y_n - \hat {\bar y}_n. \label{MainProcess}
\end{align}
The random variable $\hat {\bar y}_n$ appearing in the last line is an estimation of $\bar y_n$. The estimation is obtained using an estimation strategy $\Pi(\hat\varphi)$ on a state $\rho_T(\bar y_n)$. Each consecutive estimation is done on an independent copy of this family of states, one can picture a chain of independent identical probes \cite{Joye} interacting with the local oscillator.  The process has a discontinuity at times equal to integer multiples of $T$, note that at these points the value after the jump is assigned to the process, i.e. the process is right-continues.

\begin{definition}
We call a triple $\{ \rho_T(\varphi),\,\Pi(\hat\varphi),\, K_t\}$ an atomic clock.  A solution $y(t)$ of Eqs.~(\ref{EAbs}), (\ref{MainProcess}) is called a state of an atomic clock.
\end{definition}

Eq.~(\ref{EAbs}) describes the evolution in absence of synchronization, Eq.~(\ref{MainProcess}) describes jumps due to the synchronization and the corresponding adjustment of the frequency.  The latter equation defines a (sub)process $y_n$. This is a Markovian process that encodes the synchronization and hence has a distinguished role.  Let $ K(x,z,y) = Prob(y_n(T) = x, \bar{y}_n = z | y_n = y)$ be a joint probability distribution of $y_n(T), \hat{y}_n$ given the initial value $y_n$. Then the transfer matrix $A(y,y')$, associated to the process Eq.~(\ref{EAbs}) is given by 
\begin{align}
A(y,y') &= Prob(y_{n+1} = y | y_n = y')  \nonumber \\
&= \int_{-\infty}^\infty \int_{-\infty}^{\infty} \tr(\Pi(x-y) \rho(z)) K(x,z,y') \mathrm{d}x \mathrm{d}z. \label{rev:A}
\end{align}
Stationary distributions of $A$ are the focus of our study.

\begin{definition}[Stationary state of a clock]
We say that $y(t)$ describes a stationary state of an atomic clock if $y_n$ is a stationary process.
\end{definition}

A stationary state, $y(t)$, of an atomic clock is $T$ periodic, meaning that the joint probability distributions of $y(t_1),\dots,\,y(t_n)$ and $y(t_1+T),\dots,\,y(t_n+T)$ are identical. This in particular implies that the averaged error $\bar{y}_n$, Eq.~(\ref{taua}),
is then a stationary process.

We aim to study a situation when a clock time is unbiased, $\mathbb{E}[t_{clock}] = t $. This is true if and only if the relative frequency error has a zero average.
Consequently we say that a clock has an unbiased stationary state $y(t)$ if $\mathbb{E}[y(t)] = 0$ for all $t \geq 0$.

Whether a given clock has an unbiased stationary state is not a robust statement. It is sensitive to the noise $K_t$ and to the choice of estimation strategy. A natural question is under which conditions on $K_t$ and $\rho(\varphi)$ we can find an estimation strategy $\Pi(\hat\varphi)$ such that the clock has an unbiased stationary state. We do not know any general answer to that question and rather choose to assume more about the clock.

\begin{definition}[Unbiased clock]
\label{rev:unbiased}
We say that a clock is unbiased if $K_t \varphi$ is a martingale and the estimation strategy $\Pi(\hat\varphi)$ is $\zeta$-biased (with respect to the family $\rho_T(\varphi)$) with $|\zeta| < 1$. If a value of $\zeta$ is given we say that the clock is $\zeta$-unbiased.
\end{definition}

The above unbiasedness conditions on the local oscillator noise $K_t$ and the estimation strategy $\Pi(\hat\varphi)$ ensures that the error $y(t)$ remains unbiased provided that the initial error $y(0)$ is unbiased. The reverse statement is also true, in particular Lemma~\ref{lem:COU} implies that if the subspace of unbiased random variables is an invariant subspace of Eq.(\ref{MainProcess}) then the estimation strategy has to be $\zeta$-biased. The additional condition $|\zeta| < 1$ assures that the subspace is also attractive, this is a stability condition for the feedback loop.

In the following section we examine a clock without noise. Afterwards we study general properties of a stationary state of an unbiased clock.

\subsection{A clock without noise}
It is rather surprising that many important features of clock operation can be demonstrated in a case $K_t = 1$. 
The stochastic process $y(t)$ simplifies significantly. 
The relative frequency error $y(t)$ is constant in the intervals $(nT,\,(n+1)T)$ and jumps on its boundary. We recall that its value inside the interval was denoted by $y_n$ and the jump at the right side of the interval is $\hat{ \bar{y}}_n = \hat y_n$.
Eq.~(\ref{MainProcess}) takes a form
$$
y_{n+1} = y_n - \hat y_n.
$$
\comment{The Markovian process $\varphi_n$ has a simple transition map
\begin{align}
M(\varphi,\,\varphi')&:=Prob(\varphi_{n+1} = \varphi | \varphi_n = \varphi') \nonumber \\
M(\varphi,\,\varphi')&= \tr \left( \Pi(\varphi' - \varphi) \rho(\varphi') \right). \label{temp900}
\end{align}}


The clock time associated to a state $y(t)$ is given by
\begin{align}
\label{ClockTimeVar}
t_{clock} - t &= \int_0^t y(s) \nonumber \\
		    &=T \sum_{n=0}^{t/T} y_n.
\end{align} 
\comment{
and its variance can be computed using formula (\ref{missing}),
$$
\mathbb{E}[ (t_{clock} - t)^2 ] = \frac{T^2}{\omega_0^2} \sum_{n=0}^{t/T} \mathbb{E}[\varphi_n^2] + 2 \frac{T^2}{\omega_0^2}\sum_{n=0}^{t/T} \sum_{h=0}^{t/T-n} \mathbb{E}[\varphi_{n+h} \varphi_{n}].
$$
}

We claim that 
the variance of the clock time has a universal bound, although the variance of the frequency error can be arbitrary small. 
\begin{thm}[Unbiased clock without noise]
\label{thm:CWN}
Suppose that $K_t = 1$ and that $C_\zeta = \{ \rho_T(\varphi),\,\Pi_\zeta(\varphi) \}$ is a $\zeta$-unbiased clock. Let $y_\zeta(t)$ be a stationary state of the clock $C_\zeta$, then
\begin{equation}
\label{VWN}
\mathbb{E}[y_\zeta^2] \geq \mathbb{E}[\frac{1}{ F_T(y_\zeta)}] \frac{1-\zeta}{1+\zeta}.
\end{equation}
 The variance of clock time associated to $y_\zeta(t)$ satisfies a $\zeta$ independent bound,
\begin{equation}
\label{TWN}
\lim_{t \to \infty} \frac{\mathbb{E}[(t_{clock}- t )^2]}{t} \geq  T \mathbb{E}[\frac{1}{F_T(y_\zeta)}],
\end{equation}
where $F_T(\varphi)$ is the Fisher information associated to the family $\rho_T(\varphi)$.
\end{thm}

{\em Proof:}
Fix $\zeta$ and denote $y_n := y_\zeta(T n),\, n\in \mathbb{N}$. Then $y_n$ is a stationary process with zero mean and variance $\sigma^2:=\mathbb{E}[y_n^2]$. The Cramer-Rao inequality, Eq.~(\ref{CRUZ}), then implies
$$
\sigma^2 \geq (1 -\zeta)^2 \mathbb{E}[\frac{1}{F_T(y_n)}] + \zeta^2 \sigma^2.
$$
The inequality (\ref{VWN}) follows by solving for $\sigma^2$.

We claim that $y_n$ is a Markov chain with exponentially decaying correlations
$$
\mathbb{E}[y_{n+h} y_n] =\zeta^h \sigma^2,\quad |\zeta| \leq 1.
$$
 Then according to Example~\ref{exa:DOC} the variance of the clock time satisfies
$$
\lim_{t \to \infty} \frac{\mathbb{E}[(t_{clock} - t)^2]}{t}  =  T \sigma^2 \frac{1 + \zeta}{1 - \zeta}. 
$$
Plugging in inequality (\ref{VWN}) one obtains the bound (\ref{TWN}).

Exponential decay of correlations follows from the unbiasedness condition, 
$$\mathbb{E}[y_{n+h} | y_{n+h-1}] = \zeta y_{n+h-1},$$
 which by Lemma~\ref{XYZ} implies  that for $h \geq 1$, $\mathbb{E}[y_{n+h} y_n] = \zeta \mathbb{E}[ y_{n+h -1} y_{n}]$.
 \hfill $\square$
\vskip 3mm

In Section~\ref{sec:example} we will see an example where all bounds in the theorem are achieved. The moral to be taken is that there is a $1$-parameter family of clocks -- for a fixed family $\rho_T(\varphi)$ -- whose stationary states differ in autocorrelations, however giving an equally good clock time. 

We believe that the bound (\ref{TWN}) should be valid without assuming that the clock is unbiased. Instead only a certain ergodicity assumption to prevent a trivial counterexample of no synchronization\footnote{A process $y_{n+1} = y_n$ with initial conditions $y_0 = 0$.} should be made. Lets say an assumption that the transition map of the markov process $y_n$ has $1$ as an eigenvalue isolated by a gap from the rest of its spectrum. To support this conjecture we devote the remainder of this Section to a formulation and a proof of  the statement under an additional assumption that the process is detailed balance.

 
We fix an atomic clock $\{ \rho_T(\varphi),\,\Pi(\hat\varphi), 1 \}$ and study the corresponding Markov process 
 \begin{equation}
 \label{here}
 y_{n+1} = y_n - \hat y_{n}.
 \end{equation}
 Let $A(y,\,y')$ be the associated transfer matrix Eq.(\ref{rev:A}). A probability distribution $q$ of a stationary solution of Eq.~(\ref{here}) then satisfies $A q = q$. Equivalently, when $q$ is a solution of $Aq = q$ and $y_0$ a random variable with that probability distribution, then $y_n:=(A^*)^n y_0$ is a stationary solution of Eq.~(\ref{here}). By $A^*$ we denote the adjoint of $A$ corresponding to a duality between probability distributions and random variables.
 \begin{proposition}
 Let $A(y,\,y')$ be a transfer matrix associated to Eq.~(\ref{here}) and assume that $A$ is reversible with respect to a (stationary) probability distribution $q$ of zero mean. Let $y_n$ be the associated stationary state. Assume moreover that $1$ is a simple eigenvalue of $A$ that is isolated from the rest of the spectra, i.e. $\sigma(A) \setminus \{1\} \in B_R$ for some $R < 1$. Then the associated clock time satisfies a bound
 $$
\lim_{t \to \infty} \frac{\mathbb{E}[(t_{clock}- t )^2]}{t} \geq  T \frac{1}{\tilde{F}_T},
$$
where $\tilde{F}_T$ is the averaged Fisher information of $\rho(\varphi)$ with respect to a probability distribution $q$, see Eq.~(\ref{AF}).
 \end{proposition}
 {\em Proof:}
 For two real valued random variables $X,\,Y$ we define a scalar product $(X,\,Y)_q = \int X(y) Y(y) q(y) \mathrm{d} y$ and we denote the associated norm by $||\cdot||_q$. In terms of this product covariances of the stationary process $y_n$ are given by
 $$
 \mathbb{E}[y_{n+h} y_n] = ((A^*)^h y, y)_q
 $$
 where $y$ is an identity function. Denote $\zeta := (A^*y,\,y)_q/||y||_q^2$ then the Cramer-Rao inequality, Eq.~(\ref{QDDCR}), implies (like in Theorem~\ref{thm:CWN})
 \begin{equation}
 \label{lasteq}
 \mathbb{E}[y_n^2] \geq \frac{1}{\tilde{F}_T} \frac{1- \zeta}{1+\zeta}.
 \end{equation}
 
 The clock time variance can be expressed in terms of $A^*$ as 
 \begin{align*}
 \lim_{t \to \infty}\frac{(t- t_{clock})^2}{t}  &= T \left( ||y||_q^2 + 2\sum_{h=1^\infty} ((A^*)^h y,y)_q \right) \\
 						   &= T \left( 2 (\frac{1}{1 - A^*}y,\,y)_q - ||y||_q^2 \right),
 \end{align*}
 where the summability is guaranteed by $y \in \mathrm{Ran} ( A^* -1)$ and our spectral assumptions.
   Reversibility means that $A^*$ is hermitian with respect to the $(\cdot,\cdot)_q$ scalar product, in particular we have the following Cauchy-Schwartz inequality
   \begin{align*}
||y||_q^4 &\leq (\frac{1}{1 - A^*}y,y)_q((1-A^*)y,y)_q \\
	& \leq (\frac{1}{1-A^*}y,y)_q(1-\zeta)||y||_q^2.
   \end{align*}
   Plugging this inequality into the expression for the clock time and using Cramer-Rao inequality (\ref{lasteq}) leads to the inequality claimed in the proposition.
 \hfill $\square$

The precise statement of the conjecture mentioned above is that the claim of the proposition remains true without the assumption that $A$ is reversible with respect to a (stationary) probability distribution.

\subsection{An unbiased clock}
In this section we present the core of our results. We first prove that an unbiased clock has an unbiased stationary state, i.e. that Eq.~(\ref{MainProcess}) has a stationary solution. We then study properties of this stationary state, in particular we prove a bound from below for the associated clock time variance.

Throughout the section we explicitly compute several quantities related to a state of a $\zeta$-unbiased clock. The following will be used repeatedly in these calculations.
\begin{lemma}
\label{rev:lemma}
Let $y(t)$ be a state of a $\zeta$-unbiased clock. Then the following holds true.
\begin{enumerate}

\item[(a)] For any random variable $X$ measurable with respect to the sigma algebra $\Sigma_{n+1}$ generated by $\{ y(t),\, t < T(n+1) \}$ we have
$$
\mathbb{E}[\hat{\bar{y}}_n | X] = (1- \zeta) \mathbb{E}[\bar{y}_n | X],\quad \mbox{in particular} \quad \mathbb{E}[\hat{\bar{y}}_n  X] = (1- \zeta) \mathbb{E}[\bar{y}_n  X].
$$
\item[(b)] For any random variable $Y$ measurable with respect to the sigma algebra generated by $\{ y(t),\, t \leq Tn + s \}$ and $T>s' \geq s$ we have
$$
\mathbb{E}[{y}_n(s') | Y] = \mathbb{E}[y_n(s) | Y],\quad \mbox{in particular} \quad \mathbb{E}[{y}_n(s') Y] = \mathbb{E}[y_n(s) Y].
$$
\end{enumerate}
\end{lemma}
{\em Proof:} (a) The estimation $\hat{\bar{y}}_n$ depends on the past $t < (n+1)T$ only through the random variable $\bar{y}_n$. In fact, by the definition of $\zeta$-biased estimation we have $\mathbb{E}[\hat{\bar{y}}_n | \Sigma_{n+1}] = (1- \zeta) \bar{y}_n$. Hence,
$$
\mathbb{E}[\hat{\bar{y}}_n | X] = \mathbb{E} [\mathbb{E}[\hat{\bar{y}}_n|\Sigma_{n+1}] |X] = (1-\zeta) \mathbb{E}[\bar{y}_n |X].
$$
The second claim in (a) then follows by Eq.~(\ref{ConExp}).

The proof of (b) follows the same lines using the assumption that $y_n(s)$ is a martingale. \hfill $\square$

\comment{
In this section we prove that the stochastic process $y_n$ associated to an unbiased clock is a supermartingale. It would then follow by Doob's martingale convergence theorem that an unbiased clock has an unbiased stationary state. We finish the section by classifying these states and describing their properties.

Throughout the section we explicitly compute several quantities related to a state of an $\zeta$-unbiased clock. These computations are possible due to two relations
\begin{align}
\mathbb{E}[y(t) \hat{\bar y}_n] &= (1-\zeta) \mathbb{E}[y(t) \bar{y}_n], \quad t \leq nT, \label{rel1}\\
\mathbb{E}[y_n(t) y_n(s)] & = \mathbb{E}[y_n(t)^2], \quad t \leq s. \label{rel2}
\end{align}
The first equality follows from Lemma~\ref{XYZ} applied to a random variables triple $\{y(t),\,\bar y_n,\,\hat{\bar{y}}_n\}$. The second is an immediate consequence of $y_n(t)$ being a martingale.}

In the following theorem we explicitly state regularity assumptions, although they are covered by Assumption~\ref{assumption}. We believe that with this particular, somehow technical point, it would improve clarity.
\begin{thm}[Existence of a stationary state]
Let $\{\rho_T(\varphi),\,\Pi(\hat\varphi),\,K_t\}$ be an unbiased clock. Assume that there exist positive constants $C,\,a$ with $a <1$ such that for all $\varphi \in \mathbb{R}$ we have
$$
\int_{-\infty}^{\infty} (\hat\varphi - \varphi)^2 \tr(\rho_t(\varphi) \Pi(\hat\varphi)) \mathrm{d} \hat\varphi \leq C + a \varphi^2.
$$
Assume that $A(y,y')$ appearing in Eq.~(\ref{rev:A}) is continuous in both variables, and that $\mathbb{E}[(K_t y  - y)^2] <\infty$, for $t\in[0,\,T]$. Then the clock has an unbiased stationary state.
\end{thm}

{\em Proof:}  
We need to prove that the transfer matrix $A$, Eq.~(\ref{rev:A}), has a stationary probability distribution. We prove below that for any initial distribution $q(\varphi)$ with zero mean, the sequence of measures $(A^nq)(\varphi) \mathrm{d} \varphi$ is tight. Since the transfer matrix is Feller by our assumptions then the Krylov-Bogolioubov theorem implies that there exists a stationary measure $\mu$, i.e. for any interval $(a,b)$ it holds
$$
\mu((a,b)) = \int_a^b \int_{-\infty}^{\infty} A(y,y') \mathrm{d} y \mathrm{d} \mu(y').
$$
Continuity of $A$ then implies that $\mu$ is absolutely continuos with respect to the Lebesgue measure. Note that $\mu$ has zero mean because the probability distributions of zero mean are invariant under the map $A$.

It remains to prove the statement about tightness. We use a small observation that a family of measures $\mu_n,\, n \in \mathbb{N}$ on $\mathbb{R}$ is tight provided there exists a constant $C \leq \infty$ such that $\int \varphi^2 \mathrm{d} \mu_n (\varphi) < C$. Indeed in that case,
$$
\mu_n(\mathbb{R} \setminus (-\frac{\sqrt{C}}{\sqrt{\varepsilon}}, \frac{\sqrt{C}}{\sqrt{\varepsilon}})) \leq \int_{-\infty}^\infty \frac{\varepsilon}{C} \varphi^2 \mathrm{d} \mu_n(\varphi) \leq \varepsilon.
$$
Now let $y_0$ be a random variable with a probability distribution $q$  and $y_n$ the associated Markov process Eq.~(\ref{MainProcess}), then $\int  \varphi^2 (T^n q)(\varphi) \mathrm{d} \varphi = \mathbb{E}[y_n^2]$. We shall show that the RHS is bounded for large $n$.  We have 
\begin{align*}
\mathbb{E}[y_{n+1}^2] &= \mathbb{E}[((y_n(T) - \bar y_n) +(\bar y_n - \hat{\bar{y}}_n))^2] \\
				    &= \mathbb{E}[(\bar y_n - \hat{\bar{y}}_n)^2] + \mathbb{E}[(y_n(T) - \bar y_n)(y_n(T) - (1-2\zeta)\bar{y}_n],
\end{align*}
where to get the second line we expanded the square and used Lemma~\ref{rev:lemma}.(a), $\mathbb{E}[X \hat{\bar{y}}_n] = (1-\zeta) \mathbb{E}[X \bar{y}_n]$, on the mixed term. By the assumption of finite variance of $K_t \cdot$ the second term is bounded by a constant and we get
$$
\mathbb{E}[y_{n+1}^2] \leq Const + a \mathbb{E}[y_n^2].
$$
It then follows that $\mathbb{E}[y_n^2] \leq Const/(1- a)$ for $n$ large enough.
\hfill $\square$

Having established the existence of a stationary state, $y_n$, we now proceed to describe its properties. 
For a quantitative description of the local oscillator noise we use a variance, Eq.~(\ref{AllanVariance}),
$$
\sigma^2_{LO}(T):= \mathbb{E}[ (\bar y_n - y_n)^2],
$$
note that stionarity of $y_n$ implies that the LHS is $n$-independent.
We relate all other local oscillator quantities  to $\sigma^2_{LO}(T)$ with a help of noise dependent constants $\alpha, \beta$. These constants are defined on an appropriate place below.

We  start our description by a version of the Dick formula. This formula is traditionally derived and discussed in the frequency domain, however for our purposes the time domain is more natural.

\begin{proposition}
\label{Dick}
Let $y(t)$ be a stationary state of a $\zeta$-unbiased clock and denote $\sigma^2 = \mathbb{E}[y_n^2]$. Then for the clock time it holds
\begin{equation}
\label{clocktime}
\lim_{t \to \infty} \frac{(t_{clock} - t)^2}{t} = T\left(\sigma^2 \frac{1+\zeta}{1-\zeta} + \sigma^2_{LO}(T) \frac{1+\alpha + \zeta}{1-\zeta}\right),
\end{equation}
where $\alpha$ is defined through an equation
$$
\frac{1}{T} \int_0^T \mathbb{E}[(y_n(s)- y_n)^2] = \sigma^2_{LO}(T) \frac{\alpha + 2}{2}.
$$
\end{proposition} 
\begin{remark}
The parameter $\alpha$ is defined so that it would be consistent with an additive local oscillator noise whose variance $\sigma^2_{LO}(T) \sim T^\alpha$.
\end{remark}

{\em Proof of the proposition}:
According to Eq.~(\ref{LLN}) we have
$$
\lim_{t \to \infty} \frac{(t_{clock} - t)^2}{t} = T \left(\mathbb{E}[\bar{y}_n^2] + 2 \sum_{h=1}^\infty \mathbb{E}[\bar{y}_{n+h} \bar{y}_n] \right).
$$
We can express the quantities on the RHS using Lemma~\ref{rev:lemma}. For the first term we have
$$
\mathbb{E}[\bar y_n^2] = \mathbb{E}[(\bar y_n - y_n)^2] + \mathbb{E}[y_n^2] = \sigma^2_{LO}(T) + \sigma^2,
$$
in view of $\mathbb{E}[(\bar{y} - y_n) y_n] = 0$.
In the second term the correlations decay exponentially, for $h > 1$ 
\begin{align*}
\mathbb{E}[\bar{y}_{n+h} \bar{y}_{n}]    &= \mathbb{E}[y_{n+h} \bar{y}_n] \\
							  &= \mathbb{E}[(y_{n+h-1}(T) - \hat{\bar{y}}_{n+h-1}) \bar{y}_n] \\
							  &= \mathbb{E}[(y_{n+h-1}(T) - (1-\zeta)\bar{y}_{n+h-1}) \bar{y}_n] \\
								&= \zeta \mathbb{E}[\bar{y}_{n+h-1} \bar{y}_{n}],
\end{align*}
where we used used Lemma~\ref{rev:lemma}.(b) in the first and the last equality, and Lemma~\ref{rev:lemma}.(a) in the third equality.
 The $h=1$ term can be expressed as
 \begin{align*}
 \mathbb{E}[\bar y_{n+1} \bar y_{n}] &= \mathbb{E}[(y_{n}(T) - \hat{\bar y}_n) \bar y_n]  \\
 							&=\frac{1}{T} \int_0^T\mathbb{E}[y_{n}^2(s)] \mathrm{d}s - (1-\zeta) \mathbb{E}[\bar y_n^2]\\
							&= \zeta \sigma^2 + \frac{\alpha +2}{2} \sigma^2_{LO}(T) - (1-\zeta) \sigma^2_{LO}(T).
 \end{align*}
 After summing the geometric series and adding all the terms one gets Eq.~(\ref{clocktime}).
\hfill $\square$

We are ready to prove our main theorem that describes stationary states of unbiased clocks.

\begin{thm}
\label{thm:CWFRW}
Suppose that $C_\zeta = \{ \rho_T(\varphi),\,\Pi(\hat\varphi),\,K_t \}$ is an $\zeta$-unbiased clock and let $F_T(\varphi)$ be a Fisher information associated to the family $\rho_T(\varphi)$. Let $y(t)$ be a stationary state of the clock, then its variance $\sigma^2 = \mathbb{E}[y_n^2]$ satisfies an inequality
\begin{equation}
\label{GST}
\sigma^2 \geq \frac{1}{ F_T} \frac{1 -\zeta}{1+\zeta}  + \sigma_{LO}^2(T)\frac{\zeta^2 + \alpha \zeta +\beta - 1 -\alpha}{1-\zeta^2}, \\
\end{equation}
and for the associated clock time variance we have a bound
\begin{equation}
\label{CWFRW}
\lim_{t \to \infty}\frac{\mathbb{E}[(t_{clock}-t)^2]}{t} \geq T \frac{1}{F_T} + T \sigma^2_{LO}(T)\frac{\beta}{(1-\zeta)^2}, 
\end{equation}
where $\alpha$ was defined in the Proposition~\ref{Dick} and 
$$
\beta \sigma^2_{LO}(T) = \mathbb{E}[(K_Ty_n - y_n)^2].
$$
Above $1/F_T$ is a shorthand for $\mathbb{E}[1/F_T(\bar y_n)]$.
\end{thm}
{\em Proof:}
We follow proof of Theorem~\ref{thm:CWN} (the case $K_t =1$) only the details are more involved.

We have
\begin{align*}
\mathbb{E}[y_{n+1}^2] &= \mathbb{E}[(y_n(T) - \hat{\bar{y}}_n)^2] \\	
				    &=\mathbb{E}[(\bar{y}_n - \hat{\bar{y}}_n)^2] +  \mathbb{E}[(y_n(T) - \bar y_n)(y_n(T) - (1-2\zeta)\bar{y}_n)] \\
				    &=\mathbb{E}[(\bar{y}_n - \hat{\bar{y}}_n)^2] + \sigma^2_{LO}(T) (-1 - \alpha +\beta +\zeta \alpha). 
\end{align*}
To obtain the last equality we used $\mathbb{E}[(y_n(T) - y_n) y_n] = \mathbb{E}[(y_n - \bar{y}_n)y_n] = 0$, $\mathbb{E}[y_n(T) - y_n)\bar{y}_n] = (\alpha +2)/2 \sigma^2_{LO}(T)$ and $\mathbb{E}[(\bar y_n - y_n) \bar{y}_n] = \sigma^2_{LO}(T)$. All these relations are consequences of Lemma~\ref{rev:lemma}.(b).
Using the Cramer-Rao inequality (\ref{CRUZ}) on the RHS we then have
$$
\sigma^2 \geq \frac{(1-\zeta)^2}{F_T} + \zeta^2 (\sigma^2_{LO}(T) + \sigma^2) + \sigma^2_{LO}(T) (-1 - \alpha +\beta +\zeta \alpha).
$$
The first inequality of the Theorem follows by solving for $\sigma^2$. The inequality for the clock time follows by substituting Eq.~(\ref{GST}) into the formula for the clock time Eq.~(\ref{clocktime}).
\hfill $\square$

\begin{example}[Additive noise]
\label{additive}
For a square integrable function $f(s)$ on the interval $[0,T]$ a stochastic process $K_t\varphi = \varphi + \int_0^t f(s) \mathrm{d} W_s$ satisfies all the requirements of the above section. In particular a power law ansatz for $f(s)$ gives $\sigma^2_{LO}(T) = D T^\alpha$, where $D$ is a constant and $\alpha$ is a parameter consistent with that appearing in Proposition~\ref{Dick}, i.e.
$$
\frac{1}{T} \int_0^T \mathbb{E}[(K_s\varphi- \varphi)^2] = \sigma^2_{LO}(T) \frac{\alpha + 2}{2}.
$$
Parameter $\beta$ of Theorem~\ref{thm:CWFRW} is then given by
$$
\beta = \frac{1}{2}(\alpha+2)(\alpha +1).
$$

\end{example}

\section{Gaussian families}
\label{sec:example}
Here we aim to illustrate our concepts on a simple solvable example.  To easy the notation we put $T=1$. We also assume that the local oscillator noise is a Brownian motion, $K_{t}\varphi = \varphi + D W_t$ (generalization to any noise of the type of Example~\ref{additive} is straightforward).

Let $\rho(\varphi) = \ket{\psi(\varphi)}\bra{\psi(\varphi)}$ be a family of Gauss states on a real line,
$$
\psi_{\varphi}(x) = \braket{x}{\psi(\varphi)} = \frac{F^{1/4}}{(2\pi)^{1/4}} \exp \left( - \frac{F}{4} (x- \varphi)^2 \right).
$$
Our notation highlights the Fisher information. It is a matter of simple computation to find that the Fisher information, $F(\varphi)$, of $\rho(\varphi)$ is indeed constant and equal to $F$. 

We consider an estimation strategy $\Pi_\zeta(\hat\varphi)=(1-\zeta)^{-1}\Pi(\hat\varphi (1-\zeta)^{-1})$, where
$\Pi(\hat\varphi)$ is an orthogonal decomposition of  a position operator, $X$, on a line,
$$
X = \int \hat\varphi \Pi(\hat\varphi) \mathrm{d} \hat\varphi,\quad \Pi(\hat\varphi) = \delta(x - \hat\varphi).
$$
The conditional probability distribution of an estimate $\hat \varphi$ given parameter $\varphi$ is then
\begin{align*}
p(\hat \varphi | \varphi) &= \frac{1}{1-\zeta} \tr \left( \Pi( \frac{1}{1-\zeta} \hat\varphi) \rho(\varphi) \right) \\
				  &= \frac{1}{\sqrt{2 \pi}} \frac{F^{1/2}}{1-\zeta} \exp \left( - \frac{1}{2} \frac{F }{(1-\zeta)^2} (\hat \varphi - (1-\zeta) \varphi)^2 \right).
\end{align*}

We see that $p(\hat \varphi|\varphi)$ is a Gaussian kernel. For $\zeta = 0$ it is a symmetric heat kernel and hence unbiased. In general the estimator is multiple of unbiased estimator and the estimation strategy $\Pi_\zeta(\hat\varphi)$ is $\zeta$-biased. This can be also checked by direct integration of $\hat \varphi$ with respect to the kernel.

Now we consider a clock $\{\rho(\varphi), \,\Pi,\,K_t \}$ and its state $y(t)$. We claim that for such a clock the Markov chain $y_n$ is  Gaussian. We show this directly by computing the transfer map $A(y,\,y') = p(y_{n+1} = y | y_n = y')$, Eq.~(\ref{rev:A}).

This map can be computed by considering a joint probability distribution $p(\bar y_n,\,y_n(T) | y_n)$. It is a binomial Gaussian distribution with mean $\mu = (y_n,\,y_n)$ and a covariance matrix independent of $y_n$, whose elements might be computed in a standard way, for example using  Lemma~\ref{rev:lemma}.  The transfer map is then given by (recall that $y_{n+1} = y_n(T) - \hat {\bar y}_n$)
\begin{align*}
A(y, \, y') &= \int p(\hat{\bar y}_n = x -y,\,y_n(T) = x | y_{n} =y') \mathrm{d} x \\
				&=\int p(\hat{\bar{y}}_n = x -y | \bar y_n =z) p(\bar y_n =z,y_n(T) =x | y_n = y') \mathrm{d} x \mathrm{d} z.
\end{align*}
An integral of Gaussian kernels is itself a Gaussian, proving the claim that $y_n$ is a Gaussian process. 

The transition map can also be computed explicitly. One can either compute the involved Gaussian integrals or read the outcome form the computation in the proof of Theorem~\ref{thm:CWFRW}.
 Either way one arrives at
\begin{align}
\label{GM2}
A(y,\,y') &= \frac{1}{\sqrt{2 \pi s}} \exp \left(-\frac{1}{2 s^2} (y - \zeta y')^2 \right), \\
			s^2 &= \frac{(1-\zeta)^2}{F } + \zeta^2 2D + \frac{2}{3}  D (1 + \zeta - 2 \zeta^2).\nonumber
\end{align}

Gaussian states are determined by their mean, $\mu$, and variance, $\sigma^2$ . If we represent them by a column vector $(\mu,\,\sigma^2)^T$ then $A$ is an affine operation
$$
A \left(\! \begin{array}{c} \mu \\ \sigma^2 \end{array} \!\right) = \left(\! \begin{array}{c} \zeta \mu \\ \zeta^2 \sigma^2 + s^2 \end{array} \!\right).
$$  
It is then easy to determine a stationary Gaussian distribution, it has zero mean and a variance satisfying equation $\sigma^2 = \zeta^2 \sigma^2 + s^2$. This gives
$$
\sigma^2 =\frac{1 -\zeta}{1 + \zeta} \frac{1}{F} + \frac{2}{3(1 -\zeta^2)}  D (1 + \zeta + \zeta^2).
$$
This is exactly the RHS of the bound (\ref{GST}). Saturating this bound it also saturates the bound for the clock time. We summarize (to compare with Theorem~\ref{thm:CWFRW} put $\alpha=1, \beta=3$):

\begin{thm}
\label{thm:example}
Let $C_\zeta = (\rho(\varphi), \Pi, \Phi_\zeta)$ be a Gaussian clock described above. Then $C_\zeta$ possesses a Gaussian stationary state $y_\zeta(t)$ with variance given by
$$
\sigma^2 = \frac{1 -\zeta}{1 + \zeta} \frac{1}{F } + \sigma^2_{LO}(T)\frac{1 + \zeta + \zeta^2}{1 - \zeta^2}.
$$
The associated clock time has a standard diffusive behavior
$$
\lim_{t \to \infty} \frac{\mathbb{E}[(t_{clock} - t)^2]}{t} = T \frac{1}{F} + T\sigma^2_{LO}(T) \frac{3}{(1-\zeta)^2}.
$$
\end{thm}

\begin{remark}
There is a reason for saturation of bounds: In the Gaussian case the Cramer-Rao bound, Eq.~(\ref{CRUZ}), is saturated, because condition for equality in Cauchy-Schwarz in Eq.~(\ref{eq:CS}) is met. In fact this proves Theorem~\ref{thm:example} without any computation, however we believe that the explicit computations that were presented in this section complement a rather abstract approach of previous sections.
\end{remark}

\section{Optimization of the interrogation time}
\label{sec:opt}

The interrogation time $T$, kept fixed until now, is an adjustable parameter of an atomic clock. Long time stability of atomic clocks is susceptible and can be improved by optimizing this parameter \cite{}.  Here we find the optimal interrogation time within our model. Minimizing the bound (\ref{CWFRW}) with respect to $T$ gives an universal benchmark for the long time stability of atomic clocks formulated solely in terms of some constants describing the local oscillator noise and the frequency reference. For a Gaussian clock the bound is saturated and hence $T$ minimizing the bound gives the optimal interrogation time.

To compute the minimum of the RHS of Eq.~(\ref{CWFRW}) we need to fix a dependence of $F_T$ and $\sigma^2_{LO}(T)$ on $T$. We demonstrate the minimization on a Hamiltonian evolution (see Example~\ref{HEvolution}),
$$ F_T = 4 T^2 \Delta^2 E , $$
where $\Delta^2 E$ is the variance of energy. And we assume a phenomenological ansatz for the local oscillator noise,
$ \sigma^2_{LO}(T) = D T^\alpha. $
Minimizing the RHS of Eq.~(\ref{CWFRW}) is then straightforward and we find that for $\alpha >-1$ the minimum satisfies an equation
$$
\frac{1}{F_T} = (\alpha+1) \sigma^2_{LO}(T) \frac{\beta}{(1 - \zeta)^2}
$$
and the corresponding bound for the clock time variance is given by
$$
\lim_{t \to \infty} \frac{\mathbb{E}[(t_{clock} - t)^2]}{t} \geq \frac{\alpha +2}{\alpha +1} \left( \frac{1}{ 4 \Delta^2 E}\right)^{\frac{\alpha+1}{\alpha+2}} \left(\frac{\beta D (\alpha +1)}{(1-\zeta)^2} \right)^{\frac{1}{\alpha+2}}.
$$

In the white noise case $\alpha = -1$ the optimal interrogation time $T$ is infinite, and the corresponding bound on the clock time variance depends only the strength of the noise. This is a pure manifestation of the Dick effect.

With respect to the parameter $\zeta$ the clock time variance does not posses a minimizer. It is formally minimized by $\zeta = -1$, however there is no associated stationary solution as can be seen from the bound on the variance of the stationary state, Eq~(\ref{GST}), that has a blow up at this value of $\zeta$. This implies that the value of $\zeta$ needs to be chosen independently, for example by considering mixing times of the clock. 

\comment{It is interesting to apply this formula to the experimentally relevant case of system of $N$ spins and a  Hamiltonian
$$
H = \sigma^{(1)}_z + \sigma^{(2)}_z + \cdots +\sigma_{z}^{(N)},
$$
although this is out of our framework (see the next Section for the discussion of this important issue). The energy variance depends on the initial state, for a separable initial state it scales like $N$, while for an GHZ state
$$
\ket{GHZ} = \frac{1}{\sqrt{2}} (\ket{0} \ket{0} \dots \ket{0} + \ket{1} \ket{1}\dots \ket{1}), \quad \bra{GHZ} H^2 \ket{GHZ} = N^2.
$$
It follows that the time clock variance scales like $N^{(1 + \varepsilon)\frac{\alpha+1}{\alpha+2}}$, where $0 \leq \epsilon \leq 1$ is a parameter that depends on the initial state. }

\section{Outlooks}
\label{sec:out}
In this work we introduced a mathematical model of atomic clocks, see Eq.~(\ref{MainProcess}), and studied the stationary state of this model for the case of unbiased clocks, see Definition~\ref{rev:unbiased}. In particular we derived a lower bound for the atomic clock stability in terms of the Fisher information of the frequency reference and quantities characterizing the short time stability of the local oscillator. 

While our model incorporates environmental noise, and the Dick effect it fails to address the problem of phase-frequency ambiguity. To elaborate on the latter point, consider a family of states of $N$ spins, e.g.
\begin{equation}
\label{stam1}
\rho_T(\varphi) = e^{-i T \varphi H} \ket{N\, spins}\bra{N\, spins} e^{i T \varphi H}, 
\end{equation}
where $H = \sigma_z^{(1)} + \sigma_z^{(2)} + \cdots + \sigma_z^{(N)}$. Such a family is $2 \pi/T$ periodic in $\varphi$ and there cannot be any stationary state of the clock associated to this family. Indeed, the transfer matrix $A$ associated to this family of states, Eq.~(\ref{rev:A}) (for simplicity we consider the case of no local oscillator noise), inherits the $2 \pi/T$ periodicity and hence cannot posses a stationary state.

  When the initial state of the probe is given by $N$ independent copies of the same state 
$$
\rho(\varphi) = \rho^{(1)} (\varphi) \otimes \rho^{(1)}(\varphi) \otimes  \cdots \otimes \rho^{(1)}(\varphi),
$$
then in the large $N$ limit this state can be represented in the vicinity of $\varphi = 0$ by a Gaussian state with the Fisher information equal to the Fisher information of $\rho(0)$, see \cite{Guta, GillGuta}. This is an instance of the quantum central limit theorem. In particular, with respect to the estimation theory, the example in Section~\ref{sec:example} is generic in the large $N$ limit. From this point we considered in this study the limit $N \to \infty$ followed by the limit $t \to \infty$. To study the phase-frequency ambiguity we have to understand the behavior of solutions of Eq.~(\ref{MainProcess}) in the simultaneous limit $N, t \to \infty$.
 
  \medskip\noindent
{\bf Acknowledgements.} 
The author appreciates a help of N.~Crawford with various aspects of the probability theory and discussions with G.~M.~Graf, V.~Beaud and A.~Zarkh. 
My special thanks belongs to R.~Demkowicz-Dobrzanski and Y.~Avron for many discussions during visits at their institutions.

%

\end{document}